\documentclass[10pt,twocolumn,letterpaper]{article}

\usepackage[pagenumbers]{cvpr} %
\usepackage{float}
\usepackage[dvipsnames]{xcolor}

\definecolor{cvprblue}{rgb}{0.21,0.49,0.74}
\usepackage[pagebackref,breaklinks,colorlinks,citecolor=cvprblue]{hyperref}

\def\paperID{*****} %
\def\confName{CVPR}
\def\confYear{2024}

\title{GaussianHair: Hair Modeling and Rendering with Light-aware Gaussians}

\author{
Haimin Luo $^{1,4}$
\and 
Min Ouyang $^{1,3}$
\and   
Zijun Zhao $^{1,3}$ 
\and   
Suyi Jiang $^{1}$
\and   
Longwen Zhang $^{1,3}$
\and   
Qixuan Zhang $^{1,3}$
\and   
Wei Yang $^{2}$
\and   
Lan Xu $^{1*}$
\and   
Jingyi Yu $^{1*}$ 
\and 
$^{1}$ShanghaiTech University  
$^{2}$Huazhong University of Science and Technology\\
$^{3}$Deemos Technology  
$^{4}$LumiAni Technology
\and
}

\begin{document}

\twocolumn[{%
\maketitle
\begin{figure}[H]
\hsize=\textwidth %
\centering
\includegraphics[width=2\linewidth]{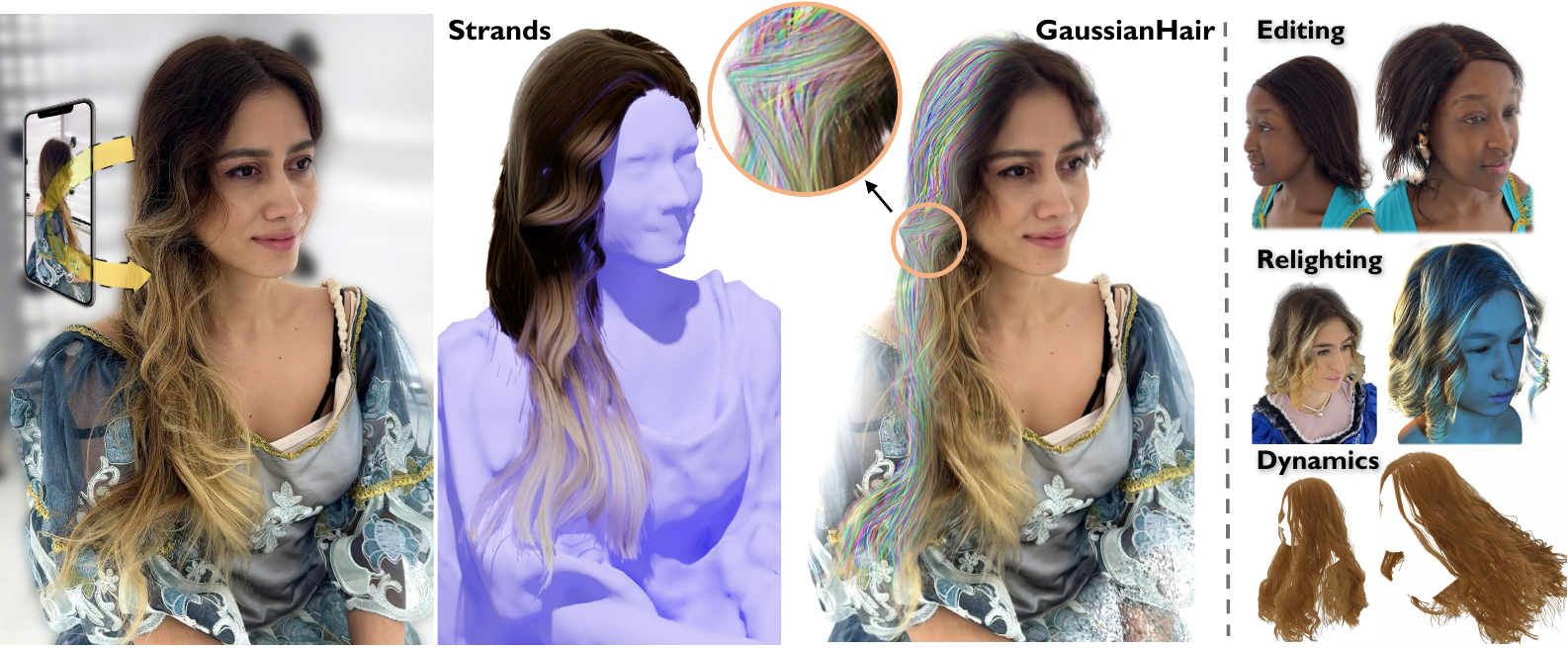}
\caption{\textbf{Illustration of Our GaussianHair}. We introduce ``GaussianHair'', a novel hair representation technique that conceptualizes a hair strand as a series of connected cylindrical 3D Gaussians. This representation facilitates the effective reconstruction of hair strands from videos captured using handheld smartphones, while also supporting an efficient scattering model. Leveraging the ``GaussianHair'' framework, image-based hair modeling extends beyond mere reconstruction, enabling advanced functionalities such as hair editing, relighting, and dynamic rendering.
  }
  \label{fig:teaser}
\end{figure}
}]

\begin{abstract}
Hairstyle reflects culture and ethnicity at first glance. 
In the digital era, various realistic human hairstyles are also critical to high-fidelity digital human assets for beauty and inclusivity.
Yet, realistic hair modeling and real-time rendering for animation is a formidable challenge due to its sheer number of strands, complicated structures of geometry, and sophisticated interaction with light. 
This paper presents GaussianHair, a novel explicit hair representation. It enables comprehensive modeling of hair geometry and appearance from images, fostering innovative illumination effects and dynamic animation capabilities.
At the heart of GaussianHair is the novel concept of representing each hair strand as a sequence of connected cylindrical 3D Gaussian primitives. This approach not only retains the hair's geometric structure and appearance but also allows for efficient rasterization onto a 2D image plane, facilitating differentiable volumetric rendering. We further enhance this model with the ``GaussianHair Scattering Model'', adept at recreating the slender structure of hair strands and accurately capturing their local diffuse color in uniform lighting. 
Through extensive experiments, we substantiate that GaussianHair achieves breakthroughs in both geometric and appearance fidelity, transcending the limitations encountered in state-of-the-art methods for hair reconstruction.
Beyond representation, GaussianHair extends to support editing, relighting, and dynamic rendering of hair, offering seamless integration with conventional CG pipeline workflows. Complementing these advancements, we have compiled an extensive dataset of real human hair, each with meticulously detailed strand geometry, to propel further research in this field.
\end{abstract}

\section{Introduction}

Hairstyles serve as visible and tangible representations of diversity in culture and ethnicity. 
For example, the intricate braiding patterns in traditional African hairstyles not only serve as a symbol of ethnic identity but also often carry cultural significance whereas the Sari Braid of India is commonly associated with traditional attire and serves as a symbol of cultural elegance. 
In the digital era, embracing a variety of hairstyles in the cyber world further helps to challenge narrow beauty standards, promote inclusivity, and contribute to a more accurate and respectful representation of human culture. Generating realistic digital versions of these hairstyles, however, remains challenging. Traditional modeling tools require extensive experience and labor to replicate the intricate details of different hair types, such as curly, straight, wavy, or coiled. Plausible hairstyle models should further conveniently support high-quality rendering to faithfully reflect their original color, texture, light penetration scattering, etc, at a high level of detail and under dynamic movements.

\begin{figure*}
	\includegraphics[width=\linewidth]{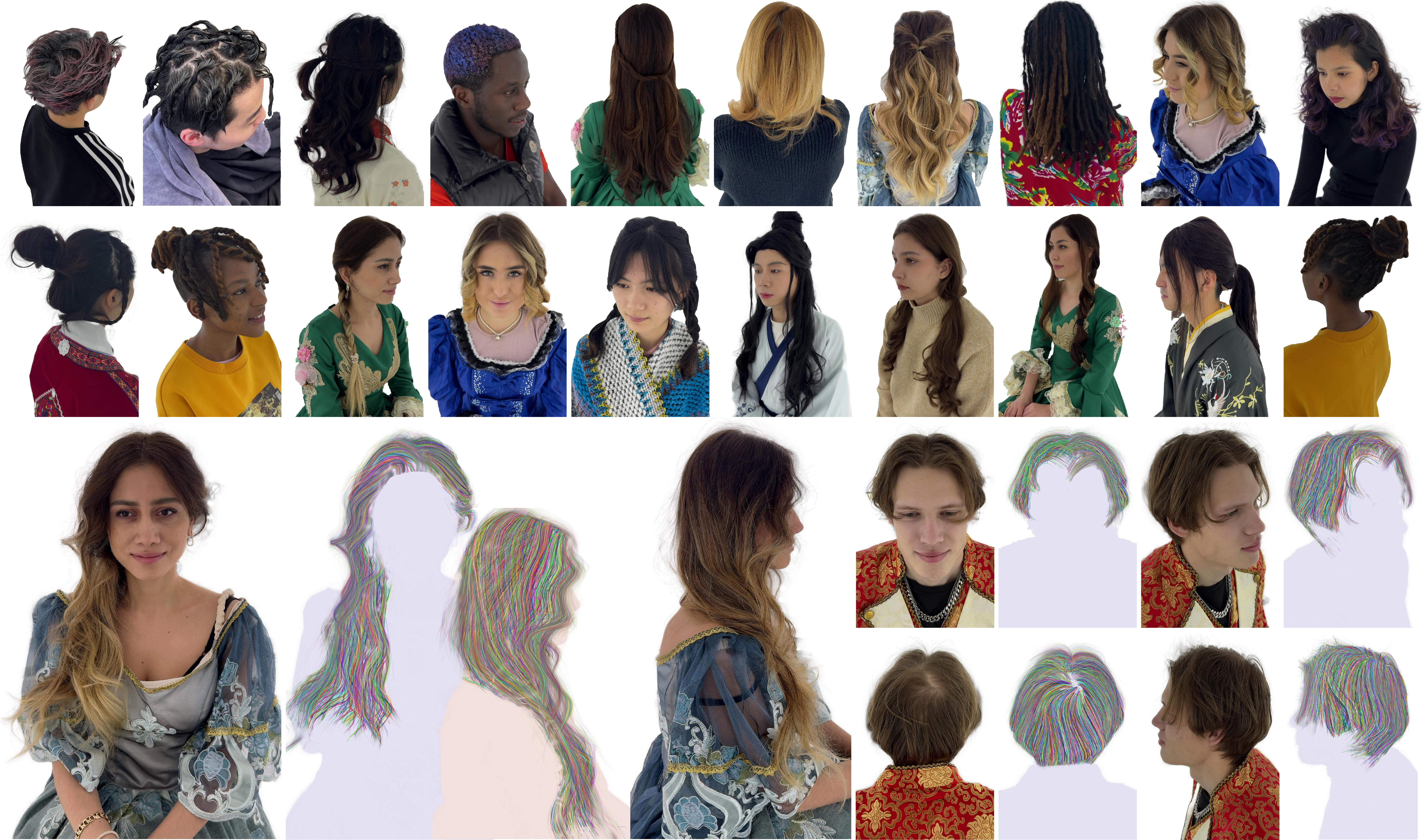}
	\caption{\textbf{Illustration of our RealHair dataset.} Our RealHair dataset represents a comprehensive and culturally diverse collection of human hairstyles, encompassing a variety of distinctive styles reflective of global hair characteristics. It comprises 281 high-resolution (4K) videos, totaling approximately 3000 frames, each meticulously annotated with detailed geometry segmentations and individual hair strand information.}
	\label{fig:datagallery}
\end{figure*}

Over the past two decades, our community has witnessed the evolution of hair modeling from explicit reconstruction methods to implicit neural representation, leveraging advances in neural modeling and rendering. Explicit geometry such as 3D polylines, or strands, has served as the dominant representation for faithful hair rendering and simulation ~\cite{rosenblum1991simulating,anjyo1992simple, Yuksel2010, blenderOrigin, Hsu2023}. These methods involved explicitly defining the geometry and properties of each individual hair strand. Building such models either relies on experienced artists or requires using expensive apparatus (e.g., with dense synchronized cameras, controlled lighting, etc) to achieve strand-accurate hair geometry and reflectance properties ~\cite{legendre2017modeling,hu2017simulation,nam2019strand,sun2021human}. They have by far been restricted to laboratory settings and cannot be readily deployed to acquire many real hairstyles in daily life. The rise of deep learning and neural networks has led to a shift towards implicit neural representation for modeling complex structures like hair. Such methods~\cite{mildenhall2021nerf, NGP, barron2021mip} unanimously aim to learn a function that directly maps input coordinates to output values without explicitly representing the underlying geometry. They are more flexible and adaptive, computationally efficient, and achieve a certain level of realism. However, implicit representations have their limitations, ranging from limited inability to produce photorealistic relighting, to difficulties in animating the hair, and to compatibility issues with traditional computer graphics (CG) pipeline workflows.

In recent advancements, hybrid representation methodologies have emerged at the forefront in the realm of image-based hair modeling. These methods ingeniously bind appearance features to three-dimensional proxies such as strands and points~\cite{rosu2022neural, Wang2020NOPC}, or to more coarse geometrical forms~\cite{sklyarova2023neural}. This process involves differentially rendering these proxies into two-dimensional features, which are then intricately decoded into hair appearance through neural rendering techniques. Notably, Neural Strands~\cite{rosu2022neural} epitomizes this approach with a strand-based generative model, learned from synthetic data. It adeptly decodes feature vectors, encoded as UV textures on the scalp, into meticulously detailed hair strands corresponding to specific scalp locations.
Following in these innovative footsteps, Neural Haircut~\cite{sklyarova2023neural} takes this strategy further by reconstructing the implicit surfaces of the head and shoulders. It then performs strand-level hair reconstruction with priors learned from synthetic datasets. However, the rendering process of these methods involves the transformation of strand representation back into feature maps, necessitating a substantial reliance on neural renderers for final appearance rendering. Such reliance inherently restricts their ability to perform per-strand animation, a key aspect in achieving dynamic and lifelike representations of hair.

In line with current research trends, we posit that an effective representation for hair modeling hinges on two key aspects: a geometry proxy model that conforms to the thin and elongated structure of hair fibers, as well as an efficient scattering and deformation model for photo-realistic hair relighting and animation. The seminal work, 3D Gaussian Splatting (3DGS) by \citet{kerbl3Dgaussians} uses a set of volumetric Gaussians with spherical harmonics appearance parameters to model 3D scenes, which brings new possibility for hair modeling. Akin to 3DGS, in this paper, we introduce \textit{GaussianHair}, a novel hair modeling and rendering scheme via light-aware Gaussian splatting. As shown in Fig.~\ref{fig:teaser}, from only video inputs using handheld devices, our GaussianHair achieves strand-level hair reconstruction and scattering effects for vivid hair rendering, editing, relighting, and dynamic animation. 

In GaussianHair, we adopt a novel primitive, dubbed cylindrical 3D Gaussian. This primitive, a 3D Gaussian with a considerably smaller radius relative to its length, allows each hair strand to be depicted as a sequence of linked cylindrical 3D Gaussian primitives, optimized through photometric supervision.
This chain of connected cylindrical Gaussians adeptly retains both the geometric structure and appearance attributes of hair, such as color and opacity, facilitating efficient rasterization onto a 2D image plane for differentiable volumetric rendering. To effectively model hair strands from images, we initialize a set of cylindrical 3D Gaussians with a fixed small radius, sampled from points on head and hair meshes, optimizing them via differentiable rasterization.
With additional guidance on 2D orientations, we derive an Oriented 3D Gaussian Field (O3GF), subsequently used to optimize geometry texture for generating coarse hair strands from a pre-trained hair strand decoder, following the approach of Neural Haircut~\cite{sklyarova2023neural}. A cylindrical 3D Gaussian is then assigned to each section between two adjacent nodes of a coarse hair strand, creating a connected sequence. We then perform further refinement to optimize the GaussianHair model.

The visual appearance of hair is significantly influenced by accurately modeling how light reflects and scatters through the hair volume. Our reconstructed GaussianHair, conforming to the thin and elongated structure of hair strands, facilitates the establishment of a scattering model. 
We employ the sophisticated UE4 approximated scattering function of the Marschner Hair Model~\cite{Marschner2003LightScattering} to simulate light scattering for the incident and exitant directions and provide an extra transmittance term to assess the light reaching each Gaussian. Our GaussianHair representation extends to support editing, relighting, and dynamic rendering of hair, offering seamless integration with conventional CG pipeline workflows, a breakthrough unseen in prior methods. Benefiting from the efficient capture setup and the strong representational ability of GaussianHair, we have compiled an extensive dataset of real human hair, each meticulously capturing detailed strand geometry, to foster further research in this field, as shown in Fig.~\ref{fig:datagallery}. We conduct extensive qualitative and quantitative experiments of GaussianHair on our diverse dataset, to demonstrate the effectiveness of GaussianHair for vivid hair modeling and rendering.

\section{Related Work} \label{sec:related_work}

Hair modeling remains challenging for computer graphics due to the geometric complexity of natural hair, its sophisticated interaction with light for realistic rendering, and the high computational demands for hair animations. Attempts to address these key issues come from both the modeling and rendering communities and the imaging and photography ones. 
\paragraph{Geometric Modeling}. Early works aim to employ some type of parametric representation of hair that provides an interface to modeling hair shape. Examples include representing hair groups as 2D parametric surfaces~\cite{koh2000real, liang2003enhanced, noble2004modelling}, 
wisps and generalized cylinders~\cite{chen1999system, yang2000cluster, xu2001v, patrick2004modelling, choe2005statistical},
multi-resolution cylinders~\cite{kim2002interactive, wang2004hair} or hair meshes~\cite{yuksel2009hair}.
Although hair shape-curving tools have become convenient, creating natural hairstyles such as the ones shown in this paper still requires tedious labor. A number of approaches have hence sought to reconstruct hair strands using computer vision techniques such as multi-view stereo (MVS). Such approaches, often referred to as image-based, include early direct reconstruction attempts ~\cite{19981351} as well as 3D volume-based solutions ~\cite{grabli2002image, paris2004capture} that generate strands in a heuristic manner. Subsequent approaches ~\cite{wei2005modeling, luo2012multi, luo2013wide, paris2008hair} have focused on using tailored acquisition apparatus, e.g., tailored imaging systems, for better reconstructing the 3D orientation field. They then triangulate the results in 3D space and grow corresponding strands. Shape primitives, e.g., ribbons, wisps, and strands, are further utilized to fit the point cloud as structure guidance\cite{bao2016realistic}. 
LPMVS~\cite{nam2019strand} further introduces a line-based PatchMatch MVS algorithm to more reliably reconstruct the point cloud alongside the actual hair strand. The point clouds are then connected into hair strands.
Based on LPMVS, ~\citet{sun2021human} adopts a per-pixel light code to further estimate the albedo color and reflectance of the captured hair for realistic rendering. 
All these approaches have focused mainly on the visible part of hair strands, i.e., hair strands at the outer regions while leaving out the interior hair volume as their reconstructions are generally noisy due to occlusions.
Several data-driven methods~\cite{hu2014robust, yu2014hybrid, zhang2017data, hu2015single} hence attempt to first build a hairstyle database that includes both the inner and outer geometry and then search for hair models from the database that best match the visible hair region. The recent deep learning-based approaches~\cite{wu2022neuralhdhair, kuang2022deepmvshair, zhou2018hairnet, saito20183d, yang2019dynamic, rosu2022neural, sklyarova2023neural} learn hair shape prior on synthetic hair dataset and can efficiently infer the entire hair volume, e.g., 3D orientation, even from a few images. 
However, these data-driven approaches rely heavily on the varieties and the quality of hairstyles of the training data. By far, public hairstyle datasets are still scarce and the only available ones \cite{karras2018progressive, karras2019style, shen2023CT2Hair, zheng2023hairstep} contain few varieties and relatively simple styles, far from the richness and diversity we observe in real life. We demonstrate that our GaussianHair representation provides a viable path to produce very high-quality hairstyles using a hand-held device. Further, we provide {a new, large hairstyle dataset  Fig.~\ref{fig:datagallery} that contains not only geometry but also appearance. 

\begin{figure*}
	\includegraphics[width=\linewidth]{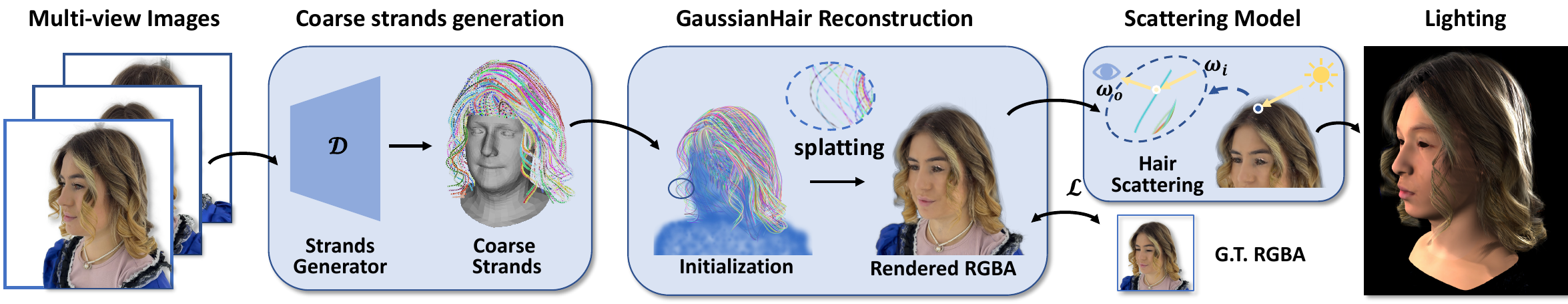}
	\caption{\textbf{Overview.} Our method employs a multi-stage process for hair modeling. Initially, an off-the-shelf decoder extracts coarse hair strands from multi-view images, which are then refined using differentiable strand-based splatting. This optimization aligns the rendered images with the ground truth. Finally, we apply a scattering model to the optimized strands, enhancing their relighting and dynamics modeling capabilities. }
	\label{fig:overview}
\end{figure*}

\paragraph{Neural Implicit Representation.} It is worth mentioning that besides explicit geometry, hairstyles can also be modeled using implicit representations. 

Such representations aim to learn continuous implicit functions from the acquired multi-view images that map spatial locations or features to some properties with Multi-Layer Perceptrons (MLPs), e.g., signed distance~\cite{park2019deepsdf, chabra2020deep, jiang2020local, wang2021neus, wang2023neus2, zhao2022human}, occupancy field~\cite{mescheder2019occupancy, peng2020convolutional, saito2019pifu, huang2020arch, he2021arch++} or radiance field ~\cite{mildenhall2021nerf, martin2021nerf, liu2020neural, bi2020neural}.
Theoretically, they can handle extremely complex geometry and complicated appearance details such as hair as they eliminate the need for modeling the topology of hairstyles. For example, ~\citet{wu2022neuralhdhair} exploits a voxel-aligned implicit function to predict 3D hair features, e.g., orientations and occupancies and further adopts an implicit hair growing scheme to obtain the complete strand model. Its concurrent work ~\citet{kuang2022deepmvshair} adopts a multi-view transformer to predict the orientation and occupancy fields. ConvNeRF~\cite{convnerf} further combines the Neural Radiance Field (NeRF) \cite{mildenhall2021nerf} and a UNet renderer for high-quality static hair geometry modeling and rendering. However, these approaches cannot readily handle re-rendering tasks such as re-lighting or hair animation. 

Potentially, subsequent variants of NeRF ~\cite{verbin2022refnerf, barron2021mip, chen2022tensorf, NGP, convnerf} can be deployed to conduct re-lighting ~\cite{bi2020neural, boss2021nerd, yao2022neilf, zhang2023neilf++, zhang2021nerfactor} or even handle dynamic scenes~\cite{peng2021neural, park2021hypernerf, cao2023hexplane}. For example, Artemis~\cite{Artemis} achieves high-quality rendering of animated animal fur driven by body movements. However, it can only handle short hair that generally does not bend or fold. Applying re-lightable NeRF to hairstyles is largely missing as most relighting tasks aim to modify low-frequency components whereas hair relighting requires as detail as specularity and scattering through individual fibers that are difficult to model using implicit representations. 

\paragraph{Volume/Point Representations.}

As a NeRF can be converted back to a volume representation, several acceleration schemes have been proposed to conduct real-time rendering, ranging from Plenoctree~\cite{yu2021plenoctrees} to the more recent Mixture of volumetric primitives (MVP)~\cite{lombardi2021mixture}. By representing scenes as a compact set of voxels., MVP mitigates the rendering artifacts caused by the limited volume resolution. Followup works ~\cite{wang2022hvh, wang2023neuwigs} extend the MVP to capturing dynamic human hair by binding voxels alongside the hair strands, achieving both hair editing and dynamic simulation. However, their rendering quality is still bound by the size of the voxels and cannot readily match photographic quality rendering. The seminal work of \citet{kerbl3Dgaussians} 3D Gaussian Splatting (3DGS) introduced Gaussian volumes as a new alternative to either mesh-based or neural-based scene representations. Its core idea of using a set of volumetric Gaussians, each with spherical harmonics appearance parameters, to model a 3D scene is revolutionary. 3DGS is also transformative as it not only achieves comparable rendering quality to NeRF and Instant-NGP \cite{NGP} at an interactive speed but more importantly it eliminates the need to use MLP neural networks and hence can be implemented on low-end graphics hardware. 3DGS has since inspired numerous subsequent extensions \cite{chen2024survey} to model human avatars, dynamic scenes, and environment-responsive models. We aim to develop a 3DGS for hair. However, the challenges are multifold. First, as hair contains tens to hundreds of thousands of fibers, the direct Gaussian representations would be prohibitively dense and large on real hairstyles. Second, even if we manage to employ 3DGS reconstruction on hair, the resulting representations are not easily animatable nor re-lightable.

We observe that 3DGS can be viewed as a special type of geometry proxy deployed in earlier image-based modeling and rendering techniques. Methods such as Light Field \cite{lightfield} and Lumigraph \cite{lumigraph} use proxies such as planes, boxes, polygonal meshes, etc, to better conduct ray interpolation. For hair strands, we set out to employ a new type of primitive, a sequence of linked cylindrical 3D Gaussian, as proxies. A major advantage of this representation is that one can easily animate hair strands and potentially conduct physical-based rendering. Our work is also inspired by the latest approaches that incorporate points~\cite{rosu2022neural, Wang2020NOPC} or meshes~\cite{sklyarova2023neural} as geometry primitives and convolutional renderers for appearance modeling. Such methods enable geometry editing but cannot modify appearance since the renderer only "memorizes" the baked hair appearance. Most related to ours, ~\citet{adaptiveshells2023} directly exploits adaptive shells to bind the volumetric region for real-time rendering and companioned animation and simulation based on mesh. The key difference is they still adopt the implicit function to represent the short hair/fur, and thus can not conduct per-strand animation and simulation.

\begin{figure}[t]
  \includegraphics[width=\linewidth]{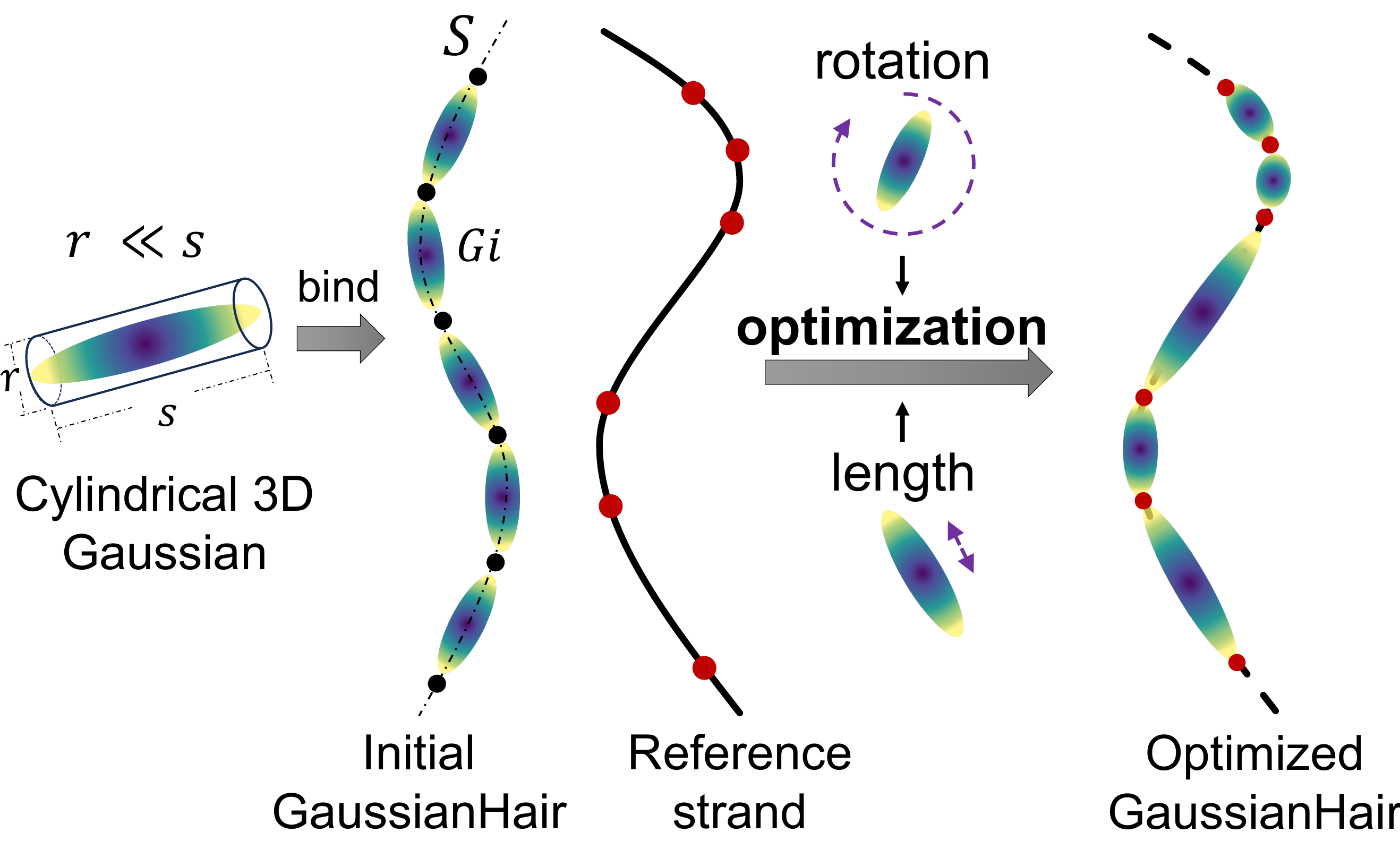}
  \caption{\textbf{Gaussian Hair Representation.} A hair strand is represented as a sequence of linked cylindrical 3D Gaussian primitives with their length $s$ significantly larger than their diameter $r$. During the modeling process, initialized strands are optimized to the optimal structures.}
  \label{fig:hair_representation}
\end{figure}

\section{GaussianHair Representation}

Traditional human hair modeling for realistic animation, requiring detailed artist input to define each strand's geometry and properties, is labor-intensive. Image-based hair modeling, an automated alternative, reconstructs hair from images or videos. However, this method typically relies on complex and costly setups like dense camera arrays with controlled lighting~\cite{wei2005modeling, luo2012multi, luo2013wide, paris2008hair}, and struggles with photorealistic rendering due to its use of explicit geometric representations like 3D polylines or parametric 3D curves.
Recent advances in neural implicit representations, such as NeRF, MipNeRF, and InstantNGP~\cite{mildenhall2021nerf, barron2021mip, NGP}, demonstrate remarkable proficiency in photorealistic hair modeling but lack the explicit geometry necessary for animation. Hybrid methodologies~\cite{rosu2022neural, sklyarova2023neural, wang2022hvh, wang2023neuwigs} blend explicit geometry with voxels for animation, yet their coarse primitives hinder accurate hair lighting effects. Furthermore, these methods often require specialized neural renderers, developed for specific hair models during training~\cite{rosu2022neural, sklyarova2023neural}. Although recent 3D Gaussian Splatting (3DGS) advancements~\cite{kerbl3Dgaussians} enable detailed scene reconstruction, their primitive structures are not naturally suited for hair strand modeling.

In response, we propose GaussianHair, an explicit volumetric hair representation optimized for both geometry and appearance. This approach leverages splatting rendering strategies to facilitate accurate hair modeling from images, allowing for dynamic animation and novel illumination.
GaussianHair represents each hair strand as a sequence of linked cylindrical 3D Gaussian primitives, optimized in position, orientation, and length through photometric supervision, as illustrated in Fig.~\ref{fig:hair_representation}. We parameterize these primitives using directional vectors and scale factors. This series preserves hair's geometric and appearance attributes and supports efficient rasterization for volumetric rendering. Critically, GaussianHair extracts explicit hair geometry from images, addressing a significant gap in current research. To aid future work, we've compiled a dataset of authentic human hair, including 281 videos with about 3000 frames, each featuring detailed strand geometry.

\paragraph{Cylindrical Gaussian Hair Representation}
We define a hair strand $\mathcal{S}$ explicitly as a sequence of cylindrical 3D Gaussian primitives, represented as $\mathcal{S} = \{ G_i \}_{i=1}^{L}$, where each $G_i$ corresponds to a segment of the hair strand. The individual cylindrical Gaussian $G_i$ is characterized by its center position $\mu$ and covariance matrix $\Sigma$. The Gaussian function for a segment is given by:
\begin{equation}
G_i(x) = e^{-\frac{1}{2}(x - \mu)^T \Sigma^{-1} (x - \mu)},
\end{equation}
where $x$ denotes a position in space. This formulation effectively segments the hair strand into $L$ distinct parts, with each part modeled as a cylindrical Gaussian, capturing the intricate geometrical features of the hair strand.

The covariance matrix $\Sigma_i$ for each cylindrical Gaussian is derived from the scaling matrix $S$ and rotation matrix $R$, formulated as $\Sigma_i = RSS^TR^T$. Here, $S$ is a diagonal matrix with diagonal elements $[d, d, s]$, representing the cylinder's diameter and length, respectively. To maintain the characteristic cylindrical shape, the length $s$ is set significantly larger than the diameter $r$. In our implementation, we use $r$ with a value of $1.0\text{e}^{-4}$ to ensure this disparity.
As a result, traditional hair strand parameters can be effectively integrated with our GaussianHair model. We represent each hair strand's particles using one endpoint of the Gaussian, defined as $\mathbf{p}_i = \mu_i + \frac{1}{2} s_i \mathbf{d}_i$, where $s_i$ is the length, and $\mathbf{d}_i$ is the direction of the $i$th cylindrical Gaussian along the strand. The direction $\mathbf{d}_i$ can be extracted from the last column of the rotation matrix $R_i$.
This approach allows for the accumulation of hair strand node points from the cylindrical Gaussians, starting from the root position $\mathbf{p}_0$ on the scalp. To enhance the hair appearance model, each cylindrical Gaussian is also equipped with an opacity value $\alpha$ and a set of spherical harmonics (SH) coefficients. These coefficients represent view-dependent color $c$, adding further realism to the hair's appearance and rendering.

During the rendering phase, all the cylindrical 3D Gaussians are projected onto the image space using the splatting technique. The color of each pixel is derived by accumulating the contributions from each Gaussian. This process involves transforming the 3D Gaussians into their 2D counterparts, denoted as $g$. The 2D covariance matrix for this transformation is defined as:
\begin{equation}
\Sigma^{'} = JW\Sigma W^TJ^T,
\end{equation}
In this equation, $J$ represents the Jacobian of the affine approximation of the projective transformation, and $W$ denotes the view transformation matrix. The 2D mean, $\mu^{'}$, is calculated as the projected center of the Gaussian onto the image plane. The contribution of a Gaussian to a pixel $u$ is then given as $w = g(u|\mu^{'},\Sigma^{'})\alpha$. Consequently, the rendered color $\mathcal{C}$ of pixel $u$ is accumulated as follows:
\begin{equation}
\mathcal{C} = \sum_{i \in N} T_i w_i c_i, \quad T_i = \prod_{j=1}^{i-1}(1 - w_j),
\end{equation} \label{eq:rendering function}
where $N$ is the set of Gaussians contributing to the pixel. To facilitate efficient optimization of all trainable parameters, we employ a fast tile-based differentiable rasterizer~\cite{kerbl3Dgaussians}. In our implementation, the hair roots are fixed, and we optimize the other parameters, namely $c, \alpha, R,$ and $S$.

\section{Image-based GaussianHair Modeling}

With our GaussianHair representation, we facilitate the acquisition of detailed hair geometry and appearance directly from images captured quickly using a standard phone camera, as shown in Fig.~\ref{fig:overview}. This approach circumvents the necessity for expensive and complex equipment typically used in traditional methods, such as camera arrays with controlled lighting setups. 

\paragraph{Expedite Data Capture and Preprocessing.} To demonstrate the ease and efficiency of our capture process, we detail a typical session. Utilizing the camera of a high-end mobile phone, specifically an iPhone 15 Pro Max, we capture a video of the subject's hair at 4Kx60 FPS. During the capture, subjects are instructed to remain as still as possible. We conducted the video capture in two rounds as follows: in the first round, the camera was positioned approximately 1 meter away from the subject's head, angled downward at about 30 degrees to focus on the hair's upper portion, and another with the camera oriented horizontally. In both scenarios, the camera is swept in one direction, ensuring comprehensive coverage of the subject's visible hair and upper torso for accurate data acquisition. Uniform white lighting is employed during the capture to facilitate precise texture recovery.
From the captured videos, we uniformly select 80-100 frames and calibrate the camera's intrinsic and extrinsic parameters using an off-the-shelf Structure from Motion (SfM) tool~\cite{agisoft2014agisoft, schonberger2016structure}.
For each frame, a set of Gabor filters is applied to compute a 2D orientation map $\mathcal{O}$~\cite{Sylvain2004orientationmap}. Additionally, we generate a portrait alpha matte $\mathcal{A}$ and a corresponding hair mask $\mathcal{M}$ using pre-trained matting models~\cite{kirillov2023segany, liu2023grounding}.
Furthermore, we fit a FLAME head model~\cite{FLAME:SiggraphAsia2017} and optimize a two-layer Instant-NSR~\cite{zhao2022human} model, akin to Neural Haircut~\cite{sklyarova2023neural}. This process yields separable hair and body meshes, denoted as $\mathbf{H}$ and $\mathbf{B}$ respectively, which serve as geometry priors for the subsequent reconstruction phase.

\begin{figure}[t]
  \includegraphics[width=\linewidth]{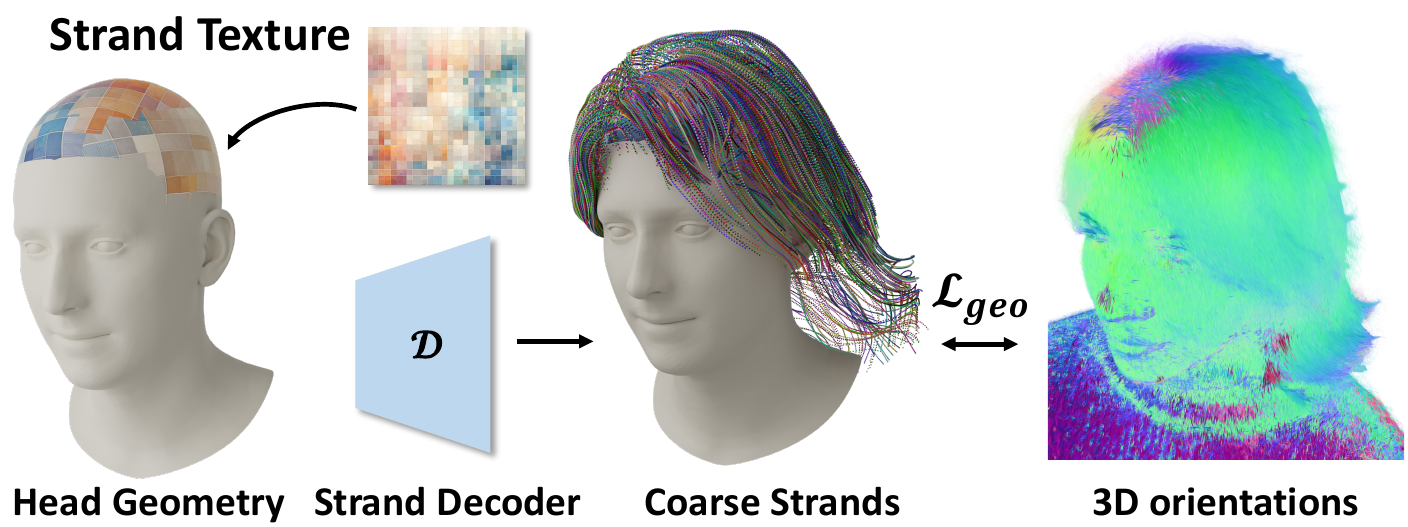}
  \caption{\textbf{Illustration of coarse strands generation.} Given fitted FLAME head mesh, we render a hair geometry texture in a differentiable manner with the 2D feature map as a UV map. Subsequently, an off-the-shelf decoder is utilized to obtain a set of hair strands which are then optimized to align with the actual hair geometry.}
  \label{fig:coarse_strands}
\end{figure}

\subsection{Oriented 3D Gaussian Field} 
The crux of our hair modeling technique lies in the precise recovery of individual hair strands from image data. Prior studies have focused on extracting directional points~\cite{rosu2022neural} or implicit orientation fields~\cite{sklyarova2023neural} solely from 2D orientation maps. 
In contrast, our approach involves generating accurate and comprehensive 3D Gaussian Orientation Fields to guide the initial creation of coarse hair strands, which are then refined in subsequent steps. We achieve this by reconstructing directional segments using additional RGB supervision with cylinder 3D Gaussians.
More specifically, we generate a set of tiny 3D Cylinder Gaussians by sampling points from the reconstructed head and hair meshes, i.e., $\mathbf{H}$ and $\mathbf{B}$ respectively. We initialize the diameter of these Gaussians, $r$, to $1.0\text{e}^{-4}$ and set the length $s$ to approximately 10 times the radius. During the optimization phase, we maintain a fixed value for $r$ across all Gaussians, while optimizing the remaining parameters through differentiable rendering, in line with the standard 3DGS procedure.
To accurately align cylinder Gaussian orientations with the natural hair strands, we render the orientations of Gaussians nearest to camera rays onto images and apply an orientation loss to minimize discrepancies, ensuring realistic hair representation.
\begin{equation}
    \mathcal{L}_{\text{ori}} = \sum_{i, j}( 1 - \mathcal{P}_i ( \{ \mathbf{d}\}, j) \cdot \mathcal{O}_i(j) ),
\end{equation}
where $i$ represents the image index and $j$ denotes a pixel within image $i$, $\mathcal{P}_i (\{ \mathbf{d} \}, j)$ refers to the projection of the set of Gaussian directions $\{ \mathbf{d} \}$ onto pixel $j$ in image $i$. $\mathcal{O}_i(j)$ is the 2D orientation at pixel $j$ in the orientation map $\mathcal{O}_i$. The symbol $\cdot$ is used to denote the dot product operation.
The total loss for optimizing our Oriented 3D Gaussian Field is defined as:
\begin{equation}
    \mathcal{L}_{\text{ogf}} = \mathcal{L}_1 + \mathcal{L}_{\text{D-SSIM}} + \mathcal{L}_{\text{ori}},
\end{equation}
where $\mathcal{L}_1 $ and $\mathcal{L}_{\text{D-SSIM}}$ are the photometric loss borrowed from the vanilla 3DGS, and all weighting factors are omitted for clarity.

Although the generated OGF is composed of oriented cylinder Gaussians, these primitives are discrete in space and do not inherently form continuous hair strands. To address this, we focus on optimizing coarse hair strands derived from the OGF, as shown in Fig.~\ref{fig:coarse_strands}. We employ a geometry texture, essentially a 2D feature map, to represent these coarse hair strands. This texture encodes information that can be decoded into lists of points along the hair strands, with their roots positioned on the scalp. The decoding process is facilitated by a Multi-Layer Perceptron (MLP) structured decoder, $\mathcal{D}$, following the methodology outlined in Neural Strands~\cite{rosu2022neural} and NeuralHirecuts~\cite{sklyarova2023neural}.
Our primary goal is to generate coarse hair strands from the Oriented Gaussian Field (OGF), represented as $\mathcal{O}^{3d}$. To facilitate this, we first refine the OGF by excluding Gaussians situated inside the head geometry $\mathbf{H}$. This step is crucial to ensure that the subsequent hair strand modeling only involves Gaussian points relevant to hair geometry.
Once the OGF is refined, we uniformly sample root points from the scalp area on the fitted FLAME head mesh $\mathbf{H}$. These root points are then used to construct the hair geometry texture $\mathbf{z}$. We decode this texture into a set of hair strand points $\mathcal{P}$, using the decoder $\mathcal{D}$. This approach, drawing upon techniques used in Neural Strands~\cite{rosu2022neural}, effectively transforms the cleaned OGF into coarse hair strands, which are crucial for the accurate and realistic representation of hair.
\begin{align}
      \mathcal{L}_{\text{geo}} = &\sum_{\mathbf{p} \in \mathcal{P}} \big ( \|\mathbf{p} - \mathbf{u}_o\|_2
       + (1 - \mathbf{d}_p\mathbf{d}_o) \big ) \Big | _{{o = NN(\mathbf{p}})} \nonumber
      \\ & + \sum_{\mathbf{o} \in \mathcal{O}^{3d}}  ( \|\mathbf{u}_o - \mathbf{p}\|_2 + (1 - \mathbf{d}_o\mathbf{d}_p) \big ) \Big | _{{ p = NN(\mathbf{o}})} ,
\end{align}
where $\mathbf{p}$ and $\mathbf{o}$ are a point in $\mathcal{P}$ and a Gaussian of $\mathcal{O}^{3 d}$ respectively, $\mathbf{o}=NN(\mathbf{p})$ means $\mathbf{o}$ is the closest point to $\mathbf{p}$ in $\mathcal{O}^{3 d}$.
Additionally, we add the hairstyle diffusion prior loss $\mathcal{L}_{dif}$ as in~\cite{sklyarova2023neural} to contain the distribution of the hair geometry texture and address the ambiguity of hair orientations, and the total loss is:
\begin{equation}
    \mathcal{L}_{\text{tex}} = \mathcal{L}_{\text{geo}}+ \mathcal{L}_{\text{dif}} ,
\end{equation}
After optimizing the geometry texture $\mathbf{z}$ with the above loss, we obtain the coarse hair stand geometries by decoding $\mathbf{z}$ using $\mathcal{D}$.

\subsection{Optimization}
To construct the GaussianHair model, we generate $N$ hair strands, each comprising $L=100$ nodes. This results in a total of $N$ fixed hair roots positioned on the head skin and $N * (L-1)$ cylinder 3D Gaussians, each containing trainable parameters including the rotation, length, color, and opacity.
We adopt the Stochastic Gradient Descent (SGD) optimizer, following the approach used in 3DGS. To ensure stability in the optimization process and to prevent large-scale oscillatory movements of the hair strands, we employ an exponential decay learning rate scheduler specifically for the rotations and scales. This strategy is crucial in achieving gradual and controlled adjustments to the hair strand geometry, thereby enhancing the realism and accuracy of the final hair model.

We adopt the original photometric loss term in 3DGS as our texture loss:
\begin{equation}
    \mathcal{L}_{\text{pho}} =  \mathcal{L}_1 + \mathcal{L}_{\text{D-SSIM}} ,
\end{equation}
We enhance our model by accumulating the opacities rendered from GaussianHair into an alpha map, denoted as $\hat{\mathcal{A}}$. To ensure the geometric accuracy of GaussianHair, we enforce consistency between $\hat{\mathcal{A}}$ and the pre-calculated alpha map $\mathcal{A}$. This supervision through alpha map consistency has been previously demonstrated to be effective in hair/fur modeling contexts, as shown in works like ConvNeRF~\cite{convnerf}:
\begin{equation}
    \mathcal{L}_{\text{alp}} = \sum_{i}|\mathcal{A}_i -\hat{ \mathcal{A}_i }| ,
\end{equation}
In addition to ensuring alpha map consistency, we also implement a smoothness regularization on the opacities of each single hair strand within the GaussianHair model. This is expressed as:
\begin{equation}
    \mathcal{L}_{\text{opa}} = \frac{1}{2} \sum_{i=0}^{N}  \sum_{j=0}^{L-1}  (|\alpha_{i, j+1} - \alpha_{i, j}| + |\hat{\alpha}_{i, j+1} - \hat{\alpha}_{i, j}|) ,
\end{equation}
where $\alpha_{i,j}$ is the opacity value of $j$th cylinder Gaussian of the $i$th strand, and $\hat{\alpha}$ is the difference in opacity of adjacent Gaussians defined by $\alpha_{i,j+1} - \alpha_{i,j}$ along the hair strand.
Besides, we apply a strand geometry regularizer to constrain the rotations and scales of adjacent Gaussians to be similar:
\begin{equation}
    \mathcal{L}_{\text{pam}} = \sum_{i=0}^{N}  \sum_{j=0}^{L-1}(|\mathbf{d}_{i, j+1} - \mathbf{d}_{i, j}| + |s_{i, j+1} - s_{i, j} | ) ,
\end{equation}
Thus the total loss function is:
\begin{equation}
    \mathcal{L}_{\text{fine}} = \mathcal{L}_{\text{pho}} + \mathcal{L}_\text{{alp}} + \mathcal{L}_{\text{opa}} + \mathcal{L}_{\text{pam}} ,
\end{equation}
During optimization, we observed hair strands mistakenly extending into face and body areas due to hair mask inaccuracies. To resolve this, we initially train head/body Gaussians, isolating the hair region. These head Gaussians are then fixed and rendered with GaussianHair in subsequent steps, allowing for accurate hair modeling without relying on hair masks, and effectively capturing areas missed by incomplete masks.

To optimize efficiency in modeling and rendering, we implement a strand-wise adaptive density control strategy, reducing the number of redundant cylinder Gaussians that have minimal impact on the final result. This involves pruning strands where a significant proportion of cylindrical Gaussians have negligible opacity, thus avoiding unnecessary computational overhead. Additionally, in regions with high gradient values, we duplicate strands to ensure adequate degrees of freedom, maintaining the integrity and detail of the hair model.

\section{GaussianHair Scattering Model} \label{PBRhair}

\begin{figure}[t]
  \includegraphics[width=\linewidth]{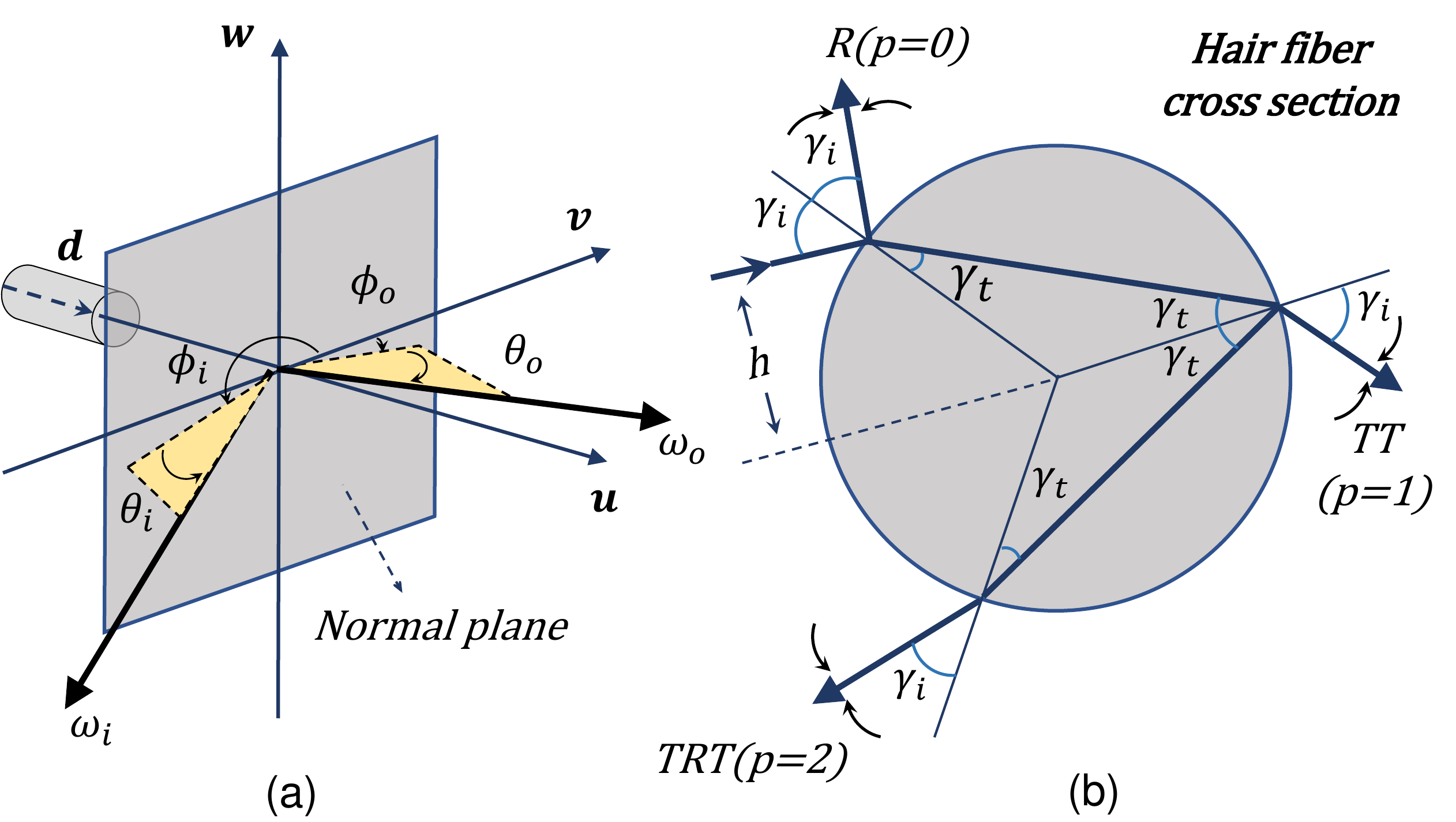}
  \caption{\textbf{Illustration of (a) hair fiber scattering geometry and (b) scattering paths.} In (a), it depicts the created coordinate system along with incident direction $\omega_i$ and extant direction $\omega_o$. In (b), three types of reflections are considered scattering through a single hair fiber, which are R (specular reflection), TT (transmission through the hair), and TRT (transmission, reflection, and transmission).}
  \label{fig:hair_bssdf}
\end{figure}

Photorealistic rendering of hair is challenging due to its complex structure and the intricate ways it interacts with light. The visual appearance of hair is significantly influenced by light's interaction, encompassing both reflection off individual strands and scattering through the hair volume. An effective hair rendering model needs to simulate these complex light interactions, addressing surface reflection, volume scattering, transparency, and shadowing, while also balancing realism with computational efficiency.
The Marschner Hair Model, as detailed in Marschner et al.\cite{Marschner2003LightScattering}, accurately represents light scattering in hair, making it a staple in film and high-end rendering for its realistic outcomes. It categorizes reflections into three types: R (specular reflection), TT (transmission), and TRT (transmission, reflection, transmission). R reflects sharply off hair strands, creating glossy highlights, while TT allows light to pass through a strand, affecting luminosity and color. TRT involves light entering a strand, undergoing internal reflection, and then exiting, yielding softer, deeper highlights. For further details, refer to Fig.\ref{fig:hair_bssdf}(b) for a visual illustration.

Our reconstructed GaussianHair accurately replicates the slender structure of hair strands and captures the local diffuse color, particularly under uniform white lighting conditions. The cornerstone of creating a light-aware rendering model lies in incorporating an effective hair scattering model. To this end, we integrate the Marschner Hair Model, which specifically addresses the scattering function of hair fibers. This function is critical in describing how the radiance from a Gaussian primitive is influenced by incident light, thereby enhancing the realism and fidelity of our hair rendering.

\subsection{Scattering Parameterization}
Hair scattering exhibits anisotropic characteristics, meaning that the way hair fibers reflect light varies significantly along their length as compared to across their width. This anisotropy arises from the elongated, cylindrical shape of hair strands, leading to directional dependence in light reflection and scattering. To quantify this, we use a scattering function $S(\omega_i, \omega_o)$, which describes the distribution of light from an incident direction $\omega_i$ to various exitant directions $\omega_o$. The color rendered when viewing from angle $\omega_o$ is essentially the cumulative result of light emittance $E(\omega_i)$ attenuated according to this scattering function. This can be mathematically represented as:
\begin{equation} \label{eq:scattering_function}
V(\omega_o) = \sum_{\omega_i} S(\omega_i, \omega_o) E(\omega_i) ,
\end{equation}
The Marschner Hair Model~\cite{Marschner2003LightScattering} articulately formulates the complex hair scattering function $S(\omega_i, \omega_o)$ for an infinitesimally small hair section. It conceptualizes this function as the sum of three components, each of which is further factorized into a longitudinal function $M(\cdot)$ and an azimuthal function $N(\cdot)$. Here, $M(\cdot)$ corresponds to the distribution along the hair's length, while $N(\cdot)$ pertains to the distribution across its width. To approximate these functions, a coordinate system is established, with the $wv$ plane perpendicular to the hair section and the $u$ axis aligned along it. This allows the longitudinal and azimuthal angles of the incident direction $\omega_i$ and the exitant direction $\omega_o$ to be defined as $(\phi_i, \theta_i)$ and $(\phi_o, \theta_o)$, respectively, as illustrated in Fig.~\ref{fig:hair_bssdf}(a). Consequently, $M(\cdot)$ and $N(\cdot)$ can be represented as functions of these angles:
\begin{equation} \label{eq:bsdf}
    S(\omega_i, \omega_o) = \sum_t M_t(\theta_h)N_t(\theta_d, \phi), \quad t\in \{R, TT, TRT\} ,
\end{equation}
where $\theta_h = (\theta_i + \theta_o) / 2$ represents the half longitudinal angle and $\theta_d = (\theta_i - \theta_o) / 2$ denotes the half difference angle. The term $t\in \{ R, TT, TRT \} $ symbolizes the three components of light scattering paths through a single hair fiber, corresponding to reflection (R), transmission-transmission (TT), and transmission-reflection-transmission (TRT), as depicted in Fig.~\ref{fig:hair_bssdf}(b).

\paragraph{GaussianHair BSDF Parameterization}

The rendering process utilizing the Marschner Model necessitates sampling points along hair strands, which involves accurately accumulating various factors including positions along the strands and the angles of light incidence and exitance. This process is inherently computation-intensive. To address this, we adopt the approximated version of the Marschner Model as implemented in the Unreal Engine 4 (UE4). This adaptation significantly enhances rendering speed while maintaining the sophistication of the original model, as discussed in Karis (2016)~\cite{Karis2016}.

In the approximation, the longitudinal scattering components are approximated through sophisticatedly designed functions that depend on the input half longitudinal angle $\theta_h$ and adjustable parameters, specifically roughness $r$ and a shift term $\beta$. The latter is particularly relevant for the second bound specular shift associated with the TRT component. This relationship is mathematically expressed as:
\begin{equation}
M_t = M_t(\theta_h; r, \beta), \quad t\in \{ R, TT, TRT \} ,
\end{equation}
Similarly, for the azimuthal reflection component, the function is determined by the reflection index $\eta$, which indicates the proportion of light reflected off the hair surface. This component for the R path is formulated as $N_{R} = N_{R}(\theta_d, \phi; \eta)$. The TT and TRT components, on the other hand, are functions of both the base color $\mathbf{b}$ of the hair fiber and the reflection index $\eta$. These functions account for the absorption of light as it travels through the hair fiber, and are represented as:
\begin{equation}
N_t = N_t(\theta_d, \phi; \eta, \mathbf{b}), \quad t \in \{ TT, TRT \} ,
\end{equation}
This approach enables a more manageable and computationally efficient simulation of hair scattering while still adhering closely to the realistic interaction of light with hair fibers.
We use the low-frequency component of the learned spherical harmonics (SH) colors as the base color $\mathbf{b}$ for each cylinder Gaussian in a hair strand. Other physical properties, such as roughness $r$, shift term $\beta$, reflection index $\eta$
are manually adjustable parameters that control various aspects of the hair's appearance.

\begin{figure}[t]
  \includegraphics[width=\linewidth]{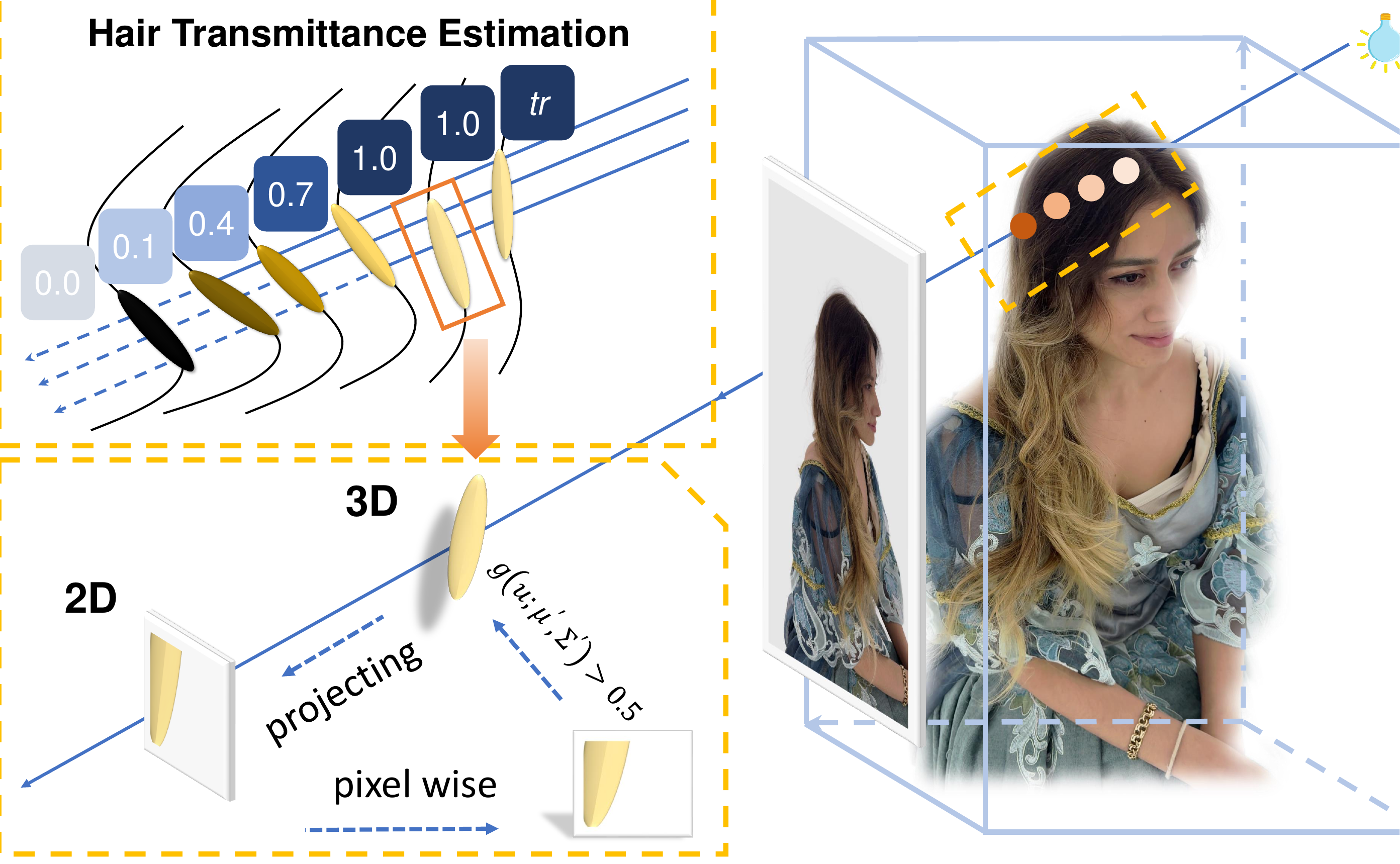}
  \caption{\textbf{Hair Transmittance Estimation.} We approximate the light transmittance $\tau$ as the inverse of the accumulated opacity $T_i$ of Gaussians from the view of the point light source. Note that for each ray, we select responsible Gaussians with contribution values greater than 0.5.}
  \label{fig:transmittance}
\end{figure}

\subsection{Multiple Scattering Approximation}
The scattering function described earlier pertains to the interaction of light with a single hair strand. However, when considering the entire hair volume, the behavior of light is more complex. Instead of undergoing a single reflection or refraction, light may bounce between multiple strands within the volume before emerging. This multiple scattering effect plays a significant role in capturing the natural appearance of hair. Modeling multi-scattering using ray tracing is computationally intensive.
Inspired by the Dual-Scattering method~\cite{dualscattering}, we approximate the multiple scattering effect by separating it into two components: the primary scattering component, characterized by light transmittance, and the local scattering component, accounting for multiple internal reflections. Specifically, we calculate how much light passing through a specific hair strand position is attenuated based on the number of hair strands it traverses. Our cylindrical Gaussian representation allows us to estimate light transmittance by considering the accumulated opacity of Gaussians during rasterization.

\begin{figure*}
	\includegraphics[width=\linewidth]{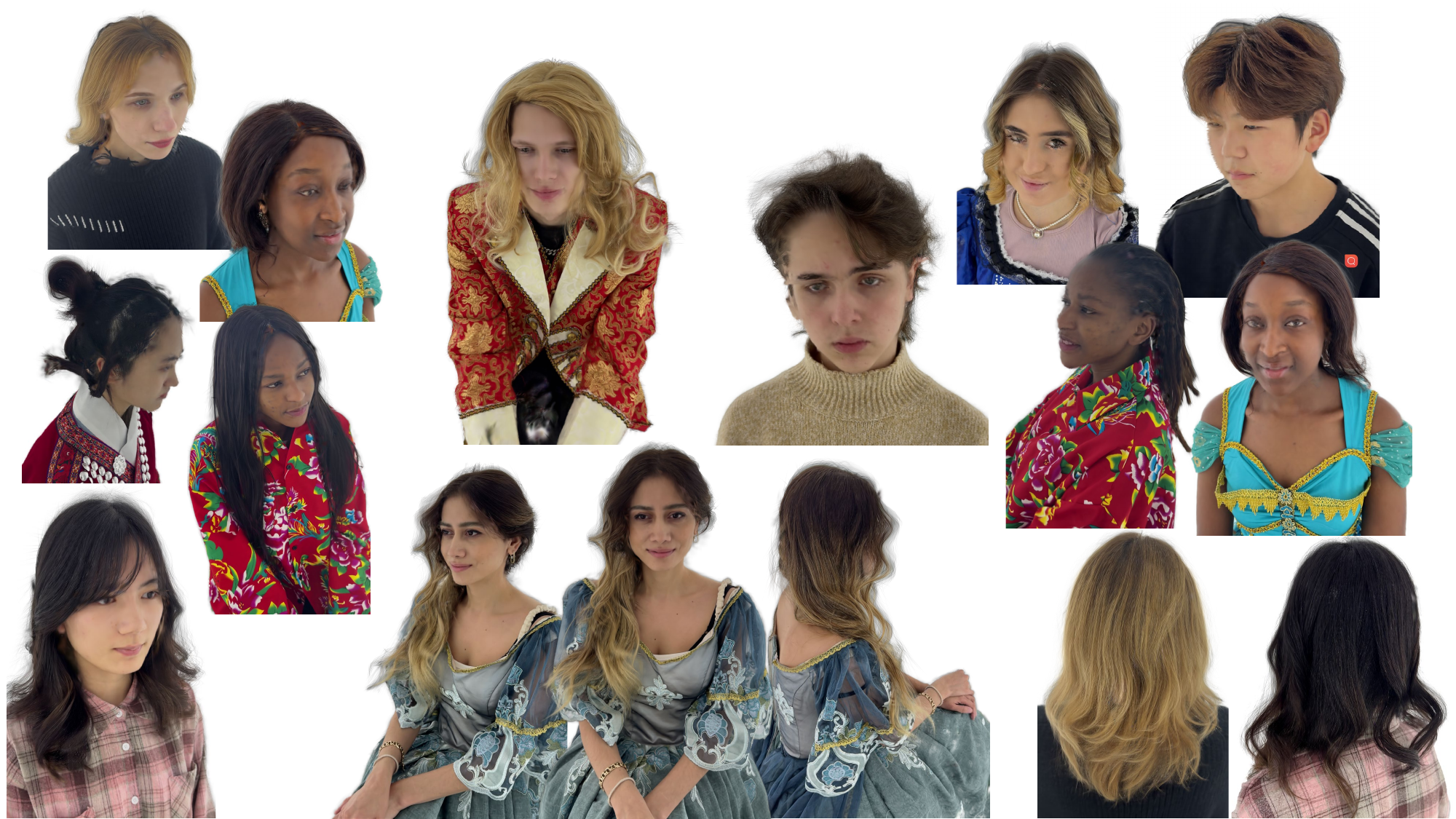}
	\caption{\textbf{Gallery of part of our rendering results.} Our method enables the reconstruction of high-fidelity hair rendering results, as demonstrated through a set of diversified hairstyles such as long, short, straight, and curly hair in our showcase. Our high-quality hair reconstruction represents a pivotal component in the advancement of the modeling of high-fidelity digital human assets.}
	\label{fig:gallery}
\end{figure*}

\paragraph{Primary Scattering} 
\label{primaryScattering}
We posit that the primary component of multi-scattering originates from the light cast by the source, and for the sake of simpler analysis, we consider a point light source illuminating the hair volume. Once we estimate the light attenuation for each hair strand, we can perform the scattering calculations using Eqn.~\ref{eq:scattering_function}.
To achieve this, we introduce an additional transmittance term denoted as $\tau$, which measures how much light has been attenuated to reach a specific Gaussian. We incorporate this term using an additional opacity rendering process.

Specifically, we only consider a point light source for simplicity. We project the 3D hair Gaussians to an image from the viewpoint of the light.
We project the 3D hair Gaussians onto an image from the viewpoint of the light source. As illustrated in Fig.~\ref{fig:transmittance}, we model the decay of transmittance $\tau$ from near to far, ranging from 1.0 to 0.0. This decay is calculated as the inverse of transmittance $T_i$ during the rasterization process, and it is assigned to the Gaussians along the ray. Specifically, we determine the responsible Gaussians along the ray based on contribution values that exceed a certain threshold, set at 0.5 in our implementation, i.e., $g(u;\mu^{'}, \Sigma^{'}) > 0.5$.

\paragraph{Local Scattering Compensation}
The primary scattering component accounts for the light that reaches the specific Gaussian after undergoing attenuation. However, it is important to note that certain portions of light, which have been scattered by other Gaussians, may also reach the same Gaussian of interest. To address this, we employ a multi-scattering approximation scheme, akin to the approach utilized in the UE4 engine, to address the absence of multiple bounce scattering.
More precisely, we interpret the light resulting from multiple bounces as a form of indirect illumination, which reaches the particular Gaussian from all directions, resembling a diffuse effect. This indirect light is subject to attenuation based on the transmittance from the light source (as described in \ref{primaryScattering}). Consequently, we model this effect as diffuse color concerning the pseudo-normal of the cylinder Gaussian:
\begin{equation}
    \mathbf{n} = \frac{\omega_o - \mathbf{d}(\mathbf{d}\cdot \omega_o)}{\|\omega_o - \mathbf{d}(\mathbf{d}\cdot \omega_o\|},
\end{equation}
where $\mathbf{d}$ is the direction of the cylinder Gaussian. $\mathbf{n}$ is the pseudo normal direction that perpendicular to $\mathbf{d}$ and in the same plane that defined by $\mathbf{d}$ and $\omega_o$. Then the local scattering component is approximated as:
\begin{equation}
    S_{local} = \sqrt{\mathbf{b}}(\frac{\mathbf{n}\cdot \omega_i + 1}{4\pi})(\frac{\mathbf{b}}{L(\mathbf{b})})^{\tau} ,
\end{equation}
where $L(\mathbf{b})$ is the relative luminance of the base color.

By combining the local scattering component with the primary scattering aforementioned, we derive the final rendering equation for each cylinder Gaussian as:
\begin{equation}
    L_o(\omega_o) = \sum_{k}^{K} \big ( S(\omega_i^k, \omega_o) + S_{local}(\omega_i^k, \omega_o)  \big ) \tau_k L_k  ,
\end{equation}
where $K$ is the number of point lights in the virtual environment. $L_k$ is the light intensity of $k$th light and $\tau_k$ is the transmittance obtained from the light pass from $k$th light.
The outgoing radiance $L_o$ is then substituted into Eq.~\ref{eq:rendering function} to obtain the final pixel color.

\section{Results} \label{sec:results}
In this section, we delineate our datasets and the experimental results. We first report the characteristics of our datasets in terms of quality, diversity, and the approaches employed for processing. We then provide comparisons with prior state-of-the-art methods and evaluation of the main technical components both qualitatively and quantitatively, followed by a detailed showcase of novel applications using our approach. Fig.~\ref{fig:gallery} shows part of our high-fidelity reconstructed hairstyles. The final subsection is devoted to limitations and discussions.

\begin{figure}[t]
  \includegraphics[width=\linewidth]{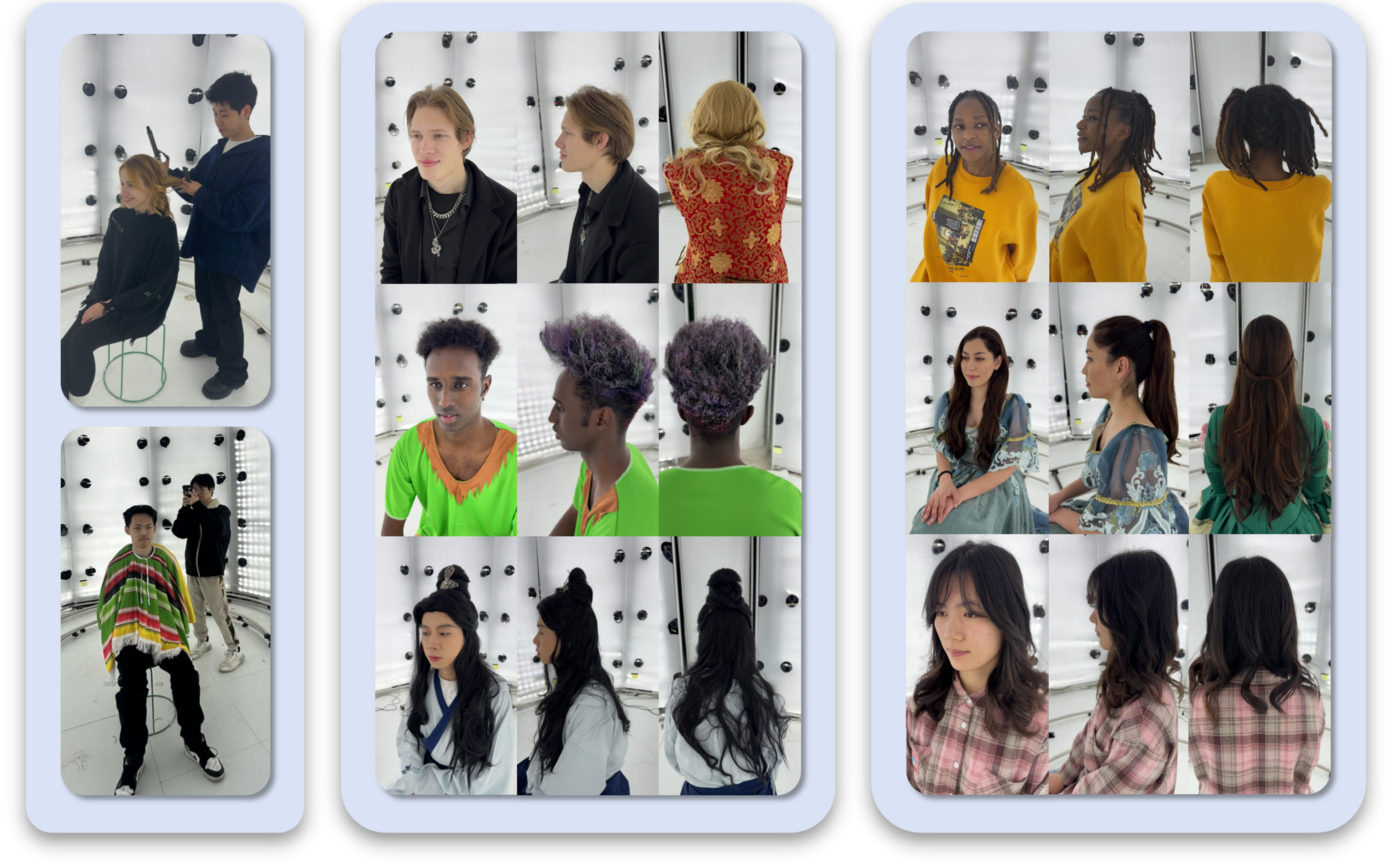}
  \caption{\textbf{Data capture}. We use a high-end mobile phone to capture surrounding videos of 4K$\times$60 FPS in a uniformly illuminated environment.}
  \label{fig:data capture}
\end{figure}

\subsection{RealHair Dataset}

To facilitate advancements in the field, we have curated an extensive dataset of authentic human hair, comprising 281 high-resolution videos, each spanning approximately 3000 frames. Fig. ~\ref{fig:data capture} delineates our sophisticated data capture methodology. Subjects were positioned centrally within our custom-built light dome, facing forward to ensure consistent, uniform illumination across their hair. High-quality video capture was conducted using advanced mobile phone cameras, capturing intricate hair dynamics and details. Then, our GaussianHair processing technique meticulously preserves the strand geometry and appearance of hair. The uniform lighting setup plays a pivotal role, enabling the captured strands in the dataset to interact with light. These enhancements not only improve the realism of digital hair but also provide seamless integration into existing CG workflows, proving invaluable for downstream applications in computer graphics and animation. 

To capture the vast spectrum of hair characteristics such as thickness, curliness, color, and sheen, our dataset embraced a global strategy by engaging models from diverse geographical regions to photograph their natural hair. To enrich the dataset, these models are outfitted in an array of custom-designed costumes during the filming process. These costumes, ranging from traditional to contemporary attire, are carefully selected to complement and accentuate the hairstyles, thus ensuring an authentic representation of various cultural backgrounds. Furthermore, to broaden the scope of our dataset, we enlist professional hairstylists to create a wide variety of unique and culturally significant hairstyles. This collaboration yields an expansive collection of hair types and styles, significantly enhancing the diversity and applicability of our dataset. To further expand our dataset concerning colored hair,  we utilize colored hair wax, allowing for the creation of hairstyles featuring a blend of colors.

Each hair data in our collection also possesses its corresponding tag. To define the hairstyles, we utilized the powerful ChatGPT-4 \cite{gpt3.5} to extract and analyze the hairstyle features in the images. Specifically, ChatGPT-4 is adopted to label the gender, approximate age range, hairstyle category, hair length range, and hair color for each image. To protect privacy and uphold ethical responsibility, we remove the geographical region information of the subjects, despite we do consider the geographical classification rules in \cite{race_stat}. We welcome the community to contribute additional samples to this dataset. This serves as a valuable contribution to honor traditional culture and broaden the aesthetic standards of the community.

\begin{figure*}
	\includegraphics[width=\linewidth]{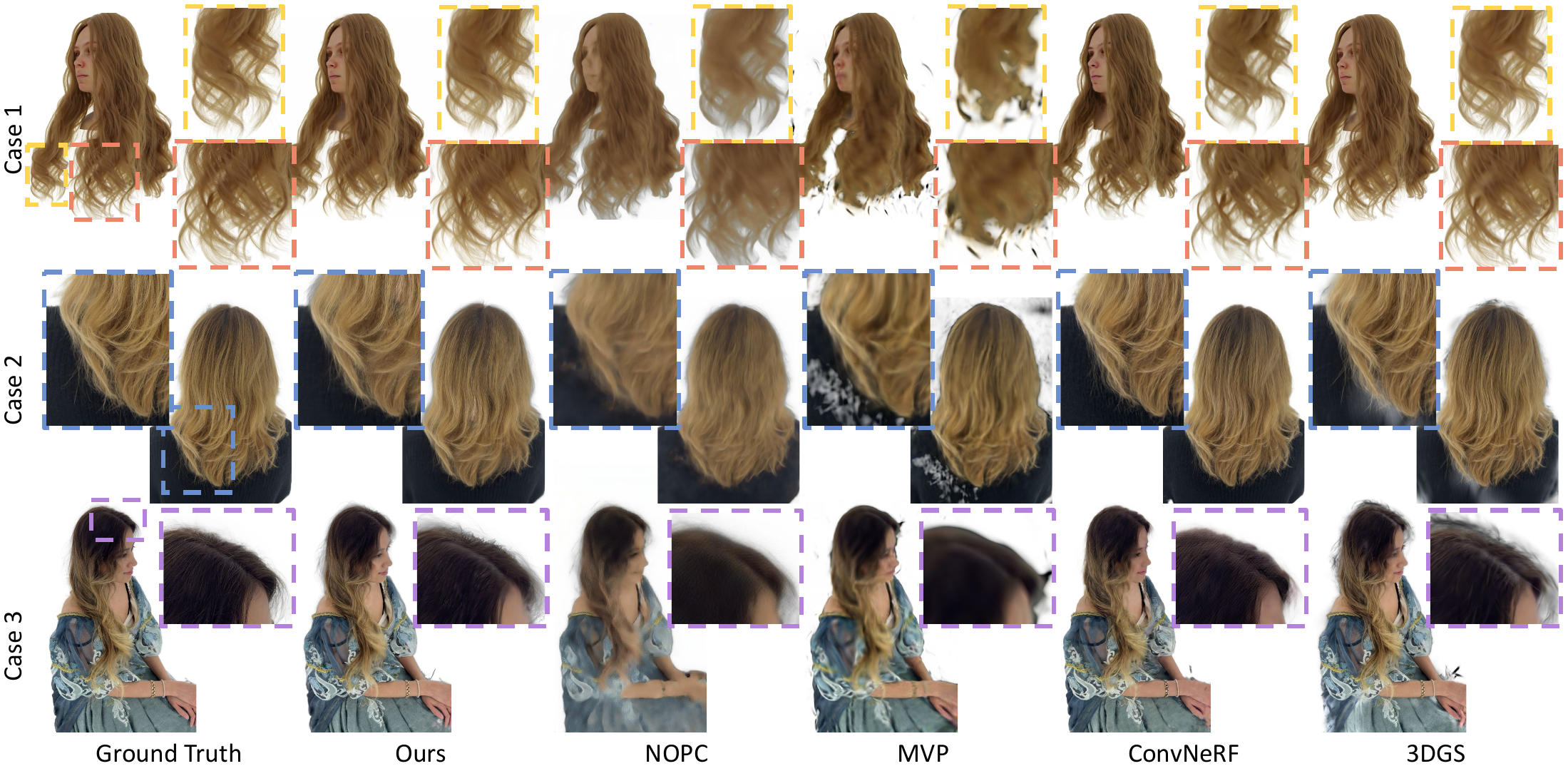}
	\caption{Qualitative Comparison between NOPC, MVP, ConvNeRF, 3D Gaussian Splatting (3DGS) and our GaussianHair. Our approach reaches a balance between reconstructing delicate hair details and achieving high rendering quality. NOPC and MVP fail to achieve high rendering results, while ConvNeRF and 3DGS have clear artifacts on hair details. Please zoom in on the image for better observation.} 
	\label{fig:comparison}
\end{figure*}

\definecolor{firstred}{rgb}{0.9, 0.2, 0.18}
\definecolor{secondblue}{rgb}{0, 0.6, 0.8}
\definecolor{firstcell}{rgb}{0.94, 0.86, 0.38}
\definecolor{secondcell}{rgb}{0.97, 0.96, 0.52}
\newcommand{\fm}{\colorbox{firstcell}}
\newcommand{\sm}{\colorbox{secondcell}}
\begin{table}[t]
\centering
\caption{\textbf{Quantitative comparison on appearance}. }
\vspace{-2ex}
\resizebox{1\linewidth}{!}{
\begin{tabular}{l|ccc|ccc|ccc}
 & \multicolumn{3}{|c}{Case 1} & \multicolumn{3}{|c}{Case 2} & \multicolumn{3}{|c}{Case 3} \\
Method & PSNR$\uparrow$ & SSIM$\uparrow$ & LPIPS$\downarrow$ & PSNR$\uparrow$ & SSIM$\uparrow$ & LPIPS$\downarrow$ & PSNR$\uparrow$ & SSIM$\uparrow$ & LPIPS$\downarrow$ \\ \hline
NOPC            & 24.66 & 0.8281 & 0.1917 & 24.79 & 0.7264 & 0.2229 & 26.61 & 0.8761 & 0.1151\\
MVP      & 22.38 & 0.8202 & 0.2548 & 23.67 & 0.7348 & 0.2773 & 26.66 & 0.8777 & 0.1248\\
ConvNeRF       & 27.65 & 0.8597 & 0.1553 & \fm{26.55} & 0.7246 & \sm{0.1817} & 27.93 & 0.8723 & 0.0958\\
3DGS      & \fm{31.97} & \fm{0.9226} & \fm{0.1212} & 25.93 & \fm{0.8406} & \fm{0.1464} & \fm{29.74} & \fm{0.9123} & \fm{0.0643}\\ 
\hline
Ours            & \sm{29.06} & \sm{0.9024} & \sm{0.1371} & \sm{26.16} & \sm{0.7550} & 0.1917 & \sm{29.54} & \sm{0.9013} & \sm{0.0767}
\end{tabular}
}
\label{tab:comparisons_appearance}
\end{table}

\subsection{Comparison}
Our comparisons with prior methods are separated into two parts: geometry comparison and appearance comparison. The geometry comparison is conducted qualitatively, adhering to two criteria: whether the geometry texture presents the thin and elongated structure of hair fibers, and whether the geometry texture visually aligns with the reference images. For appearance comparison, we select recent state-of-the-art methods that are at the forefront of hair reconstruction. The comparative analysis encompasses a spectrum of advanced techniques, ensuring a comprehensive and robust evaluation of our approach against prevailing standards.

\paragraph{Comparison on geometry.}
We compare the reconstructed geometry against the original 3DGS~\cite{kerbl3Dgaussians} and Neural Haircut~\cite{sklyarova2023neural} on our captured dataset. As illustrated in Fig.~\ref{fig:comparison_geometry}, we visualize the direction of hair strands. Specifically, for 3DGS and our results, we blend the directions of Gaussian primitives into RGB space. Hair directions in 3DGS are too unorganized and chaotic to present delicate hair structures. Neural Haircut only reconstructs the superficial layer of hair, thereby limiting the intricacy typically associated with hair structures. Our method (Ours-full) adeptly yields visually compelling results of multi-layer and intricate geometric characteristics of hair.

\paragraph{Comparison on appearance.}
We then compare the reconstructed appearance with NOPC~\cite{Wang2020NOPC}, MVP~\cite{Lombardi21}, ConvNeRF~\cite{convnerf} and the original 3DGS~\cite{kerbl3Dgaussians}. As illustrated in Fig.~\ref{fig:comparison}, NOPC and MVP fail to achieve high rendering quality and their training is inefficient. ConvNeRF recovers high-frequency details but lacks faithfulness in hair details. 3DGS tends to have blurry artifacts due to their imprecise hair geometry. As for Neural Haircut~\cite{sklyarova2023neural}, we fail to replicate comparable results with the open-source code. As shown in Tab.~\ref{tab:comparisons_appearance}, we conduct a quantitative comparison using metrics peak signal-to-noise ratio (PSNR), structural similarity index (SSIM) and Learned Perceptual Image Patch Similarity (LPIPS) with hair masks applied and our method achieves relatively high scores. ConvNeRF sometimes yields better LPIPS scores due to its use of perceptual loss. 3DGS have clear artifacts (e.g. unorganized Gaussian kernels) at the boundary region but can achieve high scores within the hair mask region. 

\begin{figure}[t]
  \includegraphics[width=\linewidth]{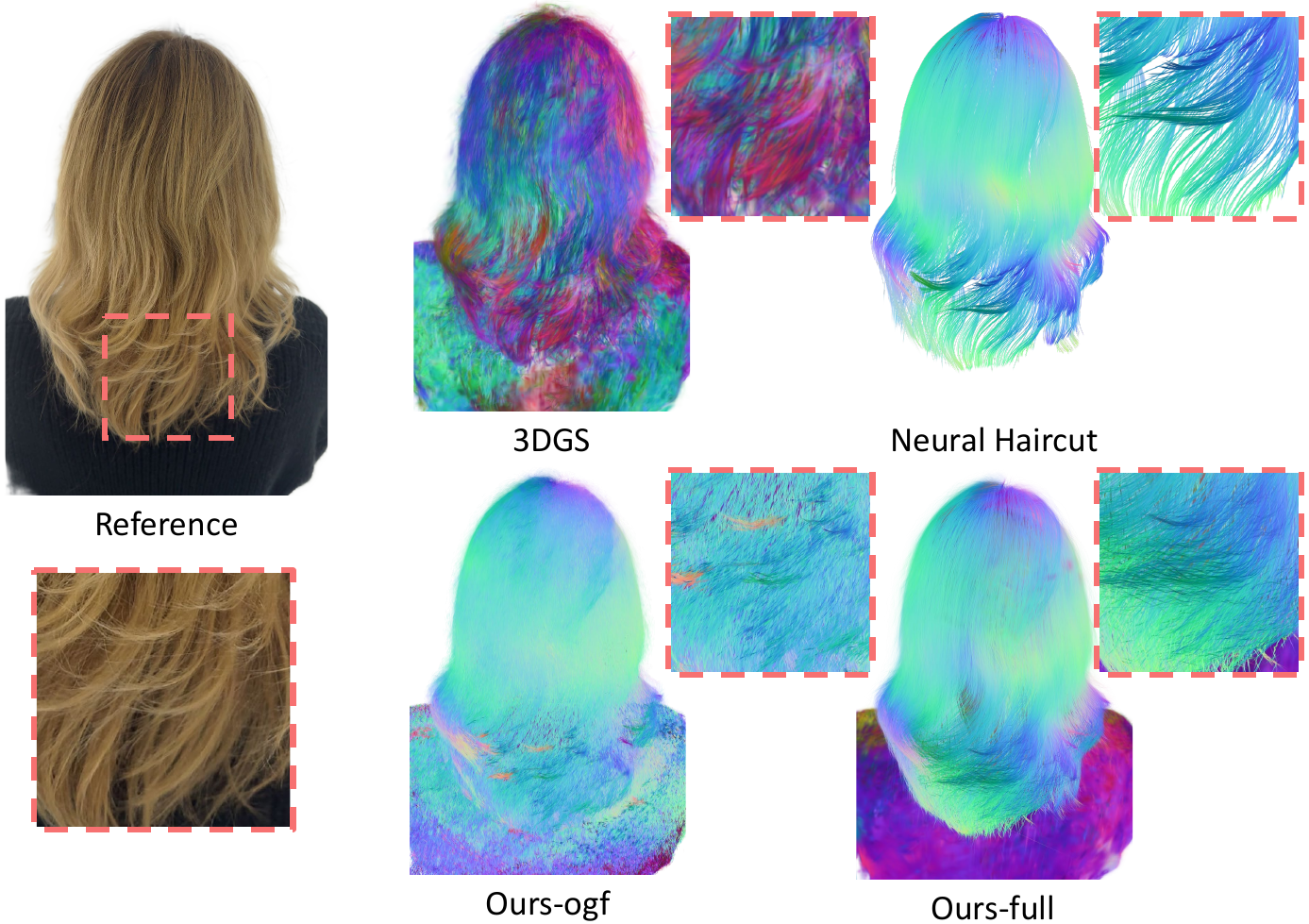}
  \caption{In the first row, we compare the geometry of 3DGS, Neural Haircut, and our method by rendering the strand direction. In the second row, we evaluate the coarse initialized geometry (Ours-ogf) and the final optimized geometry (Ours-full).}
  \label{fig:comparison_geometry}
\end{figure}

\subsection{Evaluation}

\begin{figure}[t]
  \includegraphics[width=\linewidth]{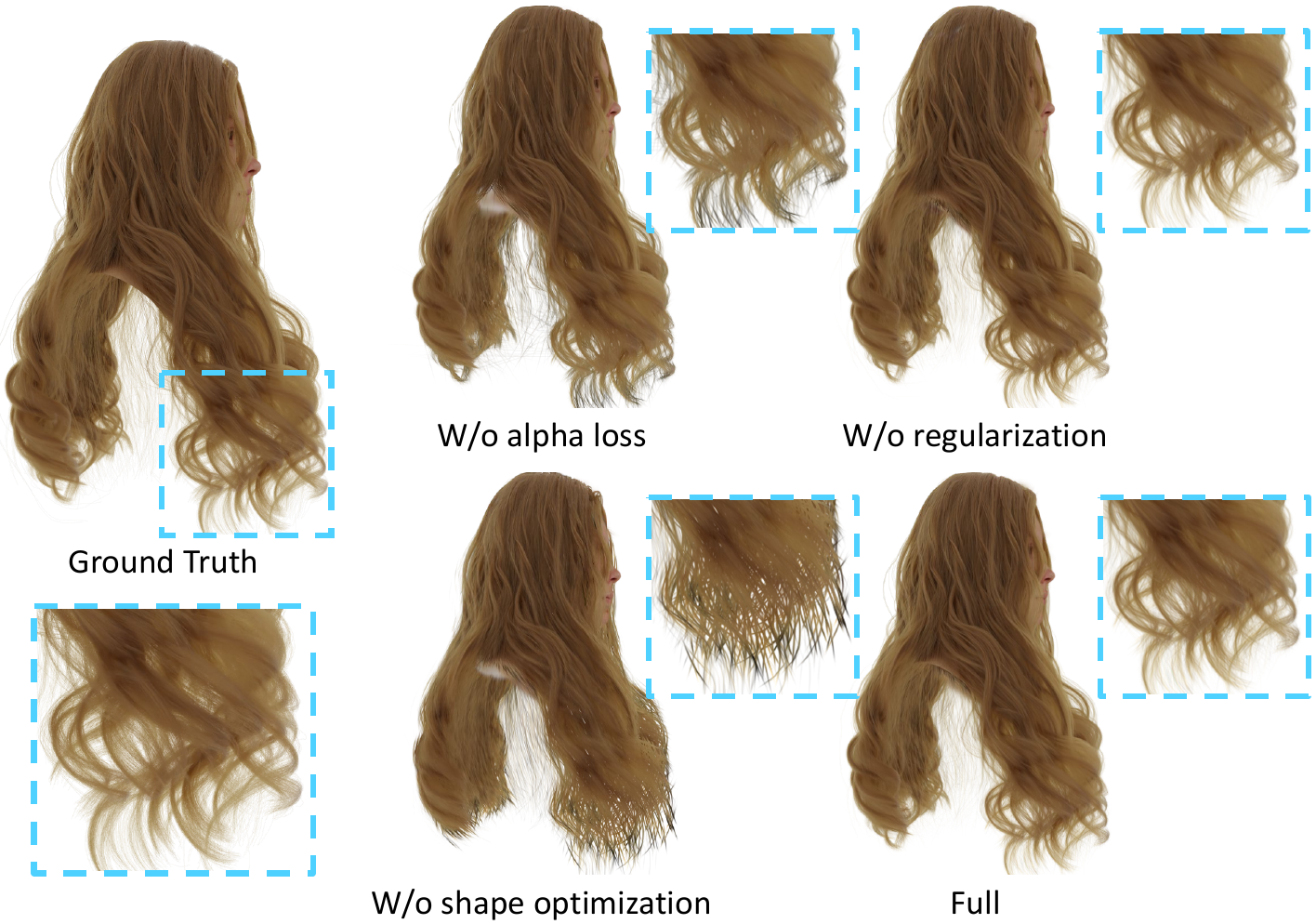}
  \caption{Qualitative evaluation of the training process. }
  \label{fig:ablation}
\end{figure}

\setlength{\tabcolsep}{2pt}
\begin{table}[t]
\centering
\caption{\textbf{Quantitative ablation on appearance.} }
\vspace{-2ex}
\resizebox{0.75\linewidth}{!}{
\begin{tabular}{l|ccc}
Method \qquad\qquad\qquad\qquad\qquad\qquad & PSNR$\uparrow$ & SSIM$\uparrow$ & LPIPS$\downarrow$ \\ \hline
w/o alpha loss ($\mathcal{L}_{alp}$)   & 25.31 & 0.8687 & 0.1570 \\
w/o regularization ($\mathcal{L}_{opa}$)      & 28.80 & 0.9013 & 0.1381 \\
w/o shape optimization      & 21.72 & 0.8075 & 0.1951 \\
\hline
full            & \textbf{29.06} & \textbf{0.9024}  & \textbf{0.1371}
\end{tabular}
}
\label{tab:ablation}
\end{table}

Here, we further evaluate the technical components of our method to validate their effect on both geometry and appearance. For geometry, we evaluate qualitatively following similar criteria mentioned in the last section. For appearance evaluation, our primary focus centers on the loss terms during the training. 

\paragraph{Ablation study on geometry.} Qualitative geometric evaluation results are illustrated in Fig.~\ref{fig:comparison_geometry} second row. \textbf{Ours-ogf} denotes the coarse strand geometry after our initialization process. Compared with our final optimized strand geometry (\textbf{Ours-full}), its hair direction is not precise, especially at the termini of hair strands.

\paragraph{Ablation study on appearance.} We evaluate our training process with qualitative and quantitative results. As illustrated in Fig.~\ref{fig:ablation}, \textbf{without alpha loss term} ($\mathcal{L}_{\text{alp}}$), notable artifacts could be observed at the termini of hair strands, which violate the fine, delicate attributes of hair. \textbf{Without smoothness regularization term} ($\mathcal{L}_{\text{opa}}$), the optimized hair strands could be observed to have a discontinuity problem, mainly caused by occlusions to various viewpoints during training. If the initialized coarse strand model is fixed \textbf{without shape optimization}, then the appearance will be fitted to the coarse geometry, resulting in significant artifacts. As illustrated in Tab.~\ref{tab:ablation}, we further adopt PSNR, SSIM, and LPIPS to evaluate the precision with ground truth images. Our full method with all components achieves the best score.

\begin{figure*}
	\includegraphics[width=\linewidth]{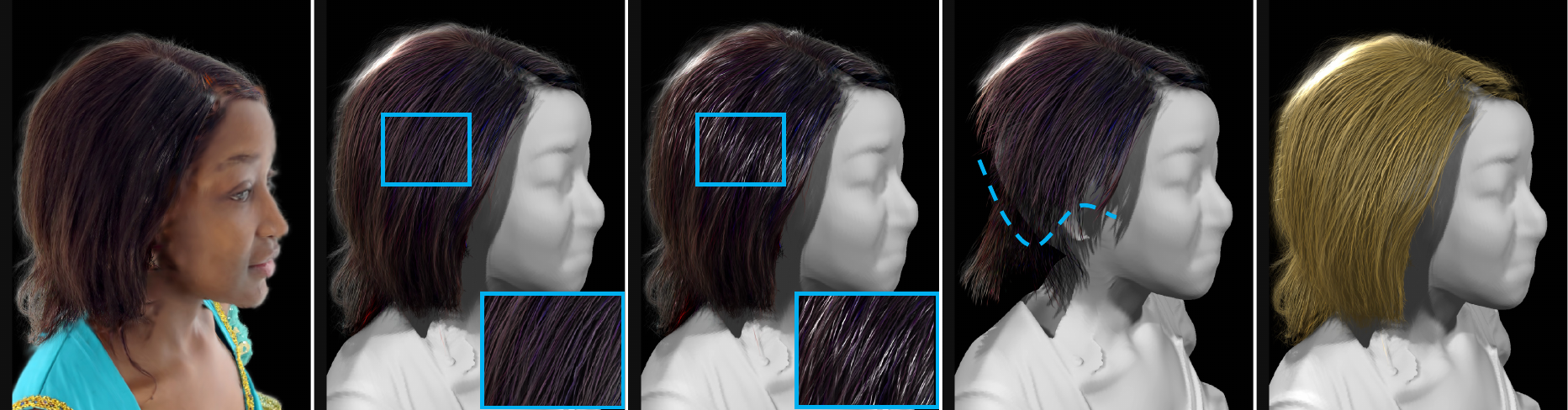}
	\caption{\textbf{Editing results.} From left to right: 1. neural rendering results 2. changed lighting 3. roughness adjustment 4. hair cutting 5. base color alteration}
	\label{fig:Editing}
\end{figure*}

\subsection{Applications}
GaussianHair already facilitates the modeling of plausible hairstyles, coupled with high-fidelity rendering. Yet, our method possesses great potential to be integrated with conventional computer graphics (CG) workflows, supporting applications such as free editing, relighting, and dynamic hair rendering. We then depict the applicability in diverse CG contexts.

\paragraph{Editing}
Once reconstructed, our method naturally facilitates a variety of editing operations, enabling the realization of a virtual salon. Thanks to our strand-based explicit representation, we can naturally adjust the length of strands and hair as illustrated in Fig.~\ref{fig:Editing} column 4. Hair color can also be changed by altering the base color as shown in column 5.
Our Gaussian scattering model brings physical attributes to the results and these attributes can be readily modified to simulate various hair textures. As shown in Fig.~\ref{fig:Editing} columns 2\&3, hairs appear to be wet after editing roughness under the same lighting condition.
These operations can significantly facilitate people in crafting their desired hairstyles.

\begin{figure*}
	\includegraphics[width=\linewidth]{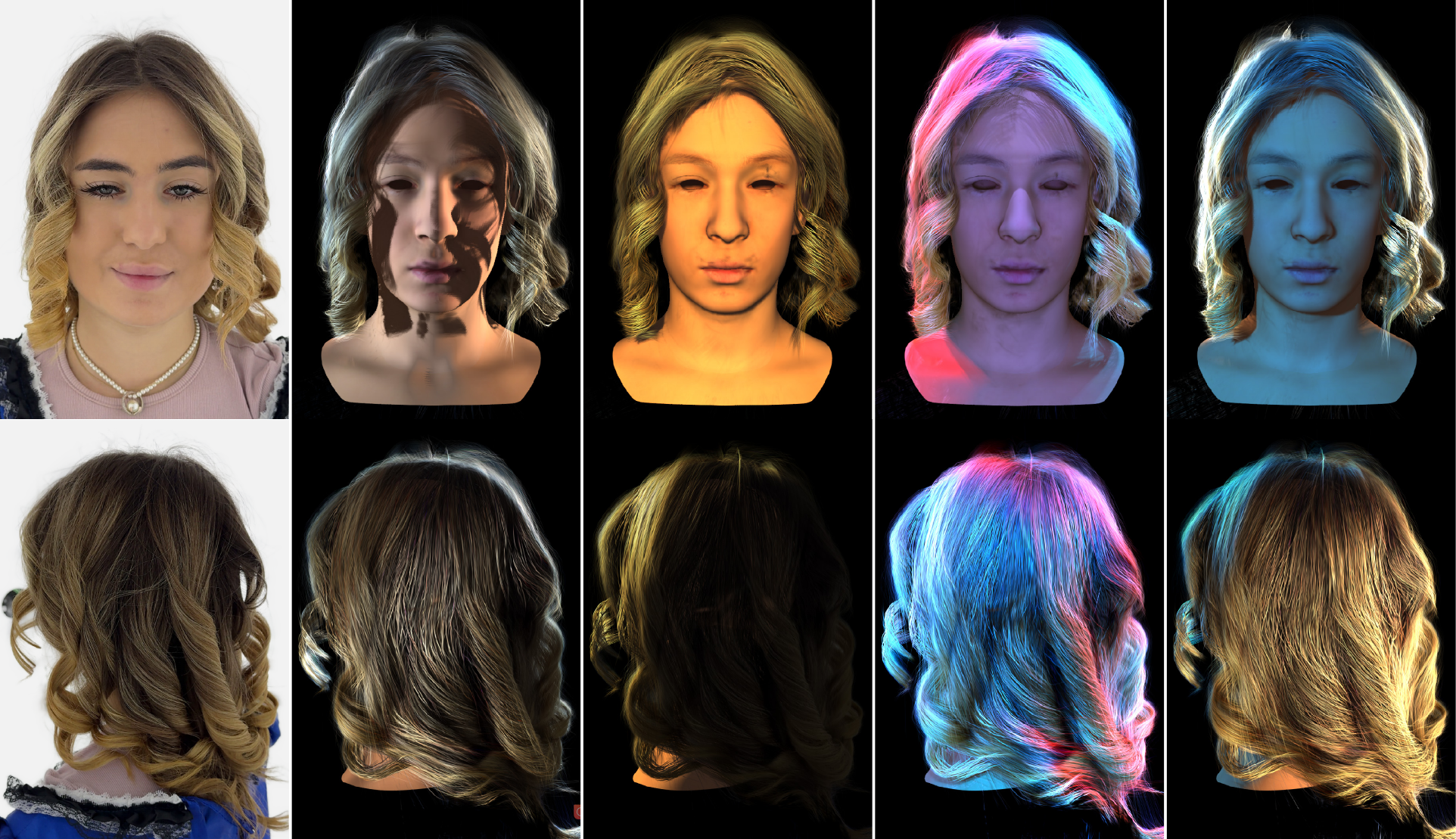}
	\caption{\textbf{Relighting.} GaussianHair can render photorealistic relighting results with various lighting conditions. Column 1 is the ground truth reference. Columns 2\&3 are rendering results under two ordinary composite lighting. Columns 4\&5 show results under Cyber style and "Avatar" style illuminations.}
	\label{fig:Relighting}
\end{figure*}

\paragraph{Relighting}
The capability to perform high-quality relighting is a critical criterion for whether a method is CG-friendly. As illustrated in Fig.~\ref{fig:Relighting}, thanks to our GaussianHair scattering model, we can obtain extremely realistic hair rendering results under various light sources. 
As shown in Fig.~\ref{fig:Relighting}, we can render with either ordinary composite lighting or film-style illuminations. Note that we combine our reconstructed hair with the face model from ChatAvatar~\cite{dreamface,chatavatar}.
This enables our method to be seamlessly integrated into the production workflow of the conventional CG pipeline for scene relighting.

\begin{figure*}
	\includegraphics[width=\linewidth]{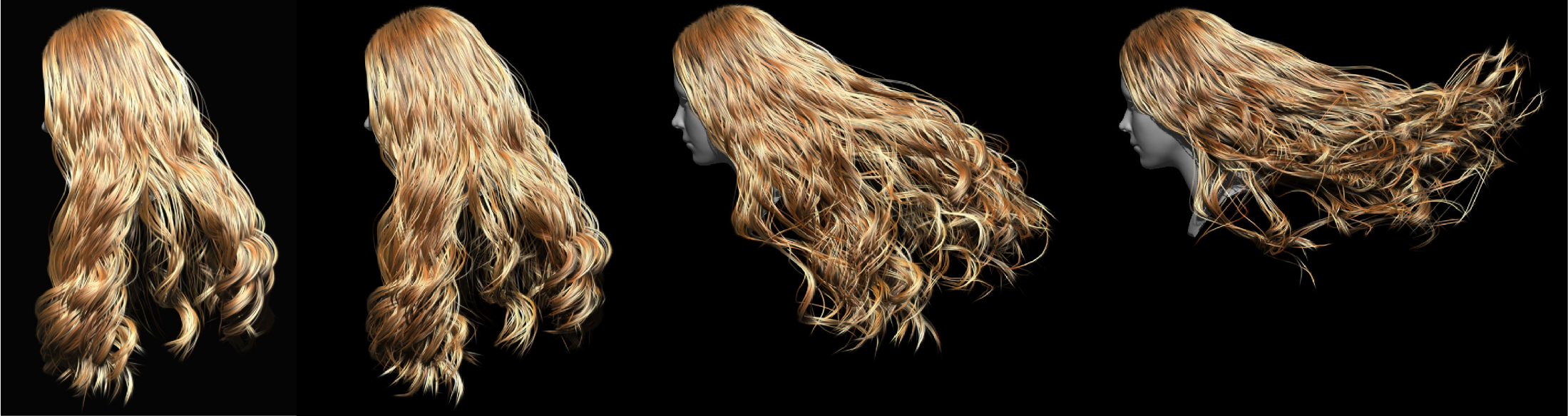}
	\caption{\textbf{Dynamic.} After importing our strand model into a conventional CG rendering engine, the returned animation result is then utilized to animate the rendered hair, simulating the effect of wind blowing.}
	\label{fig:Dynamic}
\end{figure*}

\paragraph{Dynamic Hair Rendering}
Our method is capable of dynamic hair rendering thanks to our strand-based explicit representation. By importing the reconstructed strand model into any conventional CG rendering engine, we can obtain animation results of strands which are then utilized to drive our rendering results. As shown in Fig.~\ref{fig:Dynamic}, the simulated animation of hair being uplifted by wind is utilized to drive our rendering results. 
With such capability, GaussianHair exhibits the potential to direct integration within an animation production pipeline.

\subsection{Discussion and Limitation}
\paragraph{Limitation.}
In this work, GaussianHair has demonstrated its efficacy in modeling hair strands from images, utilizing an innovative approach of initializing and optimizing cylindrical 3D Gaussians. This geometry proxy model adeptly reconstructs the nuanced, thin, and elongated structures of real hair, coupled with a Gaussian scattering model that accurately simulates light interaction. The resultant renderings exhibit impressive lighting details, showcasing the complex interplay of light and shadow on hair.
However, several limitations are identified in our approach. Firstly, the scattering model, as detailed in Sec.~\ref{PBRhair}, deviates from exact physical principles, primarily due to its reliance on an approximated BSDF parameterization from Unreal Engine and a simplified multiple scattering approach. Future improvements could involve integrating a hybrid model that leverages deep learning to enhance the accuracy of the scattering process.
Secondly, certain physical properties like roughness and the reflection index are manually adjusted within our model. Optimizing these parameters alongside the hair's albedo color through neural inverse rendering could potentially yield more accurate results.
Thirdly, there is a trade-off between rendering quality and accurate hair geometry representation in our current model. Future research will aim to refine the hair representation for improved results.
Lastly, our method faces challenges in modeling complex hairstyles, such as coiled or braided hair, due to the difficulty in reconstructing their internal structures. Future developments could explore the use of generative models to address these complex cases.

\paragraph{Ethics statement.}
Prior to the collection of our dataset, we established a protocol to ensure informed consent from all participants regarding the usage of the collected data, thereby safeguarding their portrait rights. Our selection of participants adheres to racial and cultural equality and diversity by inviting volunteers from various nationalities and cultural backgrounds. It is significant to note that our dataset is exclusively designated for research purposes, strictly prohibiting any form of commercial usage.

\section{Conclusion} \label{sec:conclusion} 

In this paper, we introduce GaussianHair, an innovative explicit representation model for human hair, which conceptualizes hair strands as a sequence of interconnected cylindrical 3D Gaussian primitives. GaussianHair excels in capturing intricate hair geometry and appearance, enabling efficient rasterization and high-quality volumetric rendering. Our model integrates the principles of the Marschner Hair Model with advancements from UE4's real-time hair rendering, culminating in the GaussianHair Scattering Model. This model is notable for its realistic rendering, surpassing current limitations in hair reconstruction.
Our experiments demonstrate the superior performance of GaussianHair in geometric and appearance fidelity, establishing a new standard in digital hair reconstruction. Beyond its representation capabilities, GaussianHair is versatile in editing, relighting, and dynamically rendering hair, seamlessly fitting into existing CG pipeline workflows.
To complement these technological advancements, we have compiled an extensive dataset of real human hair, representing a wide array of global cultures. This dataset, along with high-resolution videos and GaussianHair geometric representations, aims to promote cultural inclusivity and offer a more accurate representation of human diversity.

{
    \small
    \bibliographystyle{ieeenat_fullname}
    \bibliography{main}
}

\end{document}